\documentclass[11pt]{article}
\usepackage{jheppub}
\usepackage{amsmath,amssymb,amsfonts,graphicx,slashed,amsthm,mathtools,upgreek, enumerate, tensor, subfig}
\usepackage[dvipsnames]{xcolor}
\usepackage{arydshln}

\usepackage{comment}
\usepackage{hyperref}
\usepackage[utf8]{inputenc}
\usepackage[titletoc]{appendix}

\numberwithin{equation}{section} 
\allowdisplaybreaks

\allowdisplaybreaks

\newcommand{\be}{\begin{equation}}
\newcommand{\ee}{\end{equation}}
\newcommand{\f}{\frac}
\newcommand{\s}{\sqrt}
\newcommand{\p}{\partial}

\newcommand{\bea}{\begin{eqnarray}}
\newcommand{\eea}{\end{eqnarray}}
\newcommand{\ba}{\begin{align}}
\newcommand{\ea}{\end{align}}

\newcommand{\la}{\langle}
\newcommand{\ra}{\rangle}
\newcommand{\beq}{\begin{equation}}
\newcommand{\eeq}{\end{equation}}
\newcommand{\bra}[1]{\langle #1 |}
\newcommand{\ket}[1]{| #1 \rangle}
\newcommand{\inner}[2]{\langle #1 | #2 \rangle}
\newcommand{\avg}[1]{\langle #1 \rangle}

\DeclareMathOperator{\tr}{tr}

\title{Entanglement between two disjoint universes
}

\author[a, b]{Vijay Balasubramanian}
\author[a]{\!, Arjun Kar}
\author[c,d,a]{\!, Tomonori Ugajin}

\affiliation[\,a]{David Rittenhouse Laboratory, University of Pennsylvania,\\
209 S.33rd Street, Philadelphia, PA 19104, USA}
\affiliation[\,b]{Theoretische Natuurkunde, Vrije Universiteit Brussel (VUB), and \\ International Solvay Institutes, Pleinlaan 2, B-1050 Brussels, Belgium}
\affiliation[\,c]{Center for Gravitational Physics,
Yukawa Institute for Theoretical Physics, Kyoto University,\\
Kitashirakawa Oiwakecho, Sakyo-ku,
Kyoto 606-8502, Japan}
\affiliation[\,d]{The Hakubi Center for Advanced Research, Kyoto University,\\
Yoshida Ushinomiyacho, Sakyo-ku, Kyoto 606-8501, Japan}
\emailAdd{vijay@physics.upenn.edu}
\emailAdd{arjunkar@sas.upenn.edu}
\emailAdd{tomonori.ugajin@yukawa.kyoto-u.ac.jp}

\abstract{
We use the replica method to compute the entanglement entropy of a  universe without gravity entangled in a thermofield-double-like state with a disjoint gravitating universe. Including wormholes between replicas of the latter gives an entropy functional which includes an ``island" on the gravitating universe.  We solve the back-reaction equations when the cosmological constant is negative to show that this island coincides with a causal shadow region  that is created by the entanglement in the gravitating geometry.
At high entanglement temperatures, the island contribution to the entropy functional leads to a bound on entanglement entropy, analogous to the Page behavior of evaporating black holes. We demonstrate  that the entanglement wedge of the non-gravitating universe grows with the entanglement temperature until, eventually, the gravitating universe can be entirely reconstructed from the non-gravitating one.
}


\begin{document}

\maketitle

\parskip=10pt

\section{Introduction}
Recent work suggests a  new way in which semiclassical gravity encodes the entanglement between underlying quantum degrees of freedom.   The idea is that the  entropy of a  non-gravitating system entangled with a gravitating system can  be determined by the area of an isolated ``island" in the latter \cite{Almheiri:2019hni}.  This idea,  
inspired by  holographic entanglement entropy  \cite{Ryu:2006bv,Ryu:2006ef,Hubeny:2007xt}  and its quantum correction \cite{Faulkner:2013ana, Engelhardt:2014gca,Almheiri:2019psf, Penington:2019npb},
is realized quantitatively by a modified  replica method for computing entanglement entropy  which includes new saddlepoints of the semiclassical gravity path integral: wormholes between replicated copies of the gravitating geometry \cite{Penington:2019kki,Almheiri:2019qdq}.

These ideas have been largely developed and applied in the context of evaporating black holes in AdS space, where one imagines Hawking radiation captured in a non-gravitating ``reservoir'' just outside the spacetime boundary.  The island proposal in this context gives a way to evade unbounded increase of the entropy of Hawking radiation (which would violate unitarity \cite{Page:1993wv}), entirely in semiclassical gravity.  This is remarkable, as it would have been natural to expect that analyzing unitarity in black hole evaporation would require access to the underlying microstates and their entanglement patterns.

If these ideas are fundamental to quantum gravity, they should apply generally, for any boundary conditions.\footnote{See \cite{Anegawa:2020ezn, Hashimoto:2020cas, Gautason:2020tmk,Krishnan:2020oun,Hartman:2020swn,Dong:2020uxp} for recent discussions of asymptotically flat black holes.}  To test this, we will set up the simplest possible scenario: two disjoint but entangled two-dimensional universes, both carrying quantum field theories, and one also gravitating according to the Jackiw-Teitelboim (JT) model.  The disconnection between the two universes allows us to evade the a key technical challenge in \cite{Almheiri:2019qdq}, namely smoothly welding the gravitating and non-gravitating theories together.  We will tune the amount of entanglement between the two universes, and ask how gravity affects changes in the entanglement structure.

We begin in Sec.~\ref{sec:setup}  with two disjoint universes $A$ and $B$ with a bipartite Hilbert space $\mathcal{H}_{\text{tot}} = \mathcal{H}_{A} \otimes \mathcal{H}_{B}$.  Both universes are two dimensional and support quantum field theories, but only $B$ has gravity turned on. We will give $B$ a non-trivial classical geometry, like an eternal  black hole.  $A$ and $B$ cannot communicate classically, but we will set them up in an entangled state with a thermofield-double-like structure controlled by a parameter $\beta$.  Our goal then is to ask how the von Neumann entropy of the reduced density matrix on the non-gravitating universe $A$ is affected by the gravity in universe $B$. To calculate this entanglement entropy,  we  employ the replica trick, i.e, starting from the R\'enyi entropy ${\rm tr} \: \rho^{n} $ for all fixed integer $n$, we  take the $n \rightarrow 1$ limit of $\partial_n {\rm tr} \: \rho^n$. The R\'enyi  entropy is  computed by a path integral on $n$ replicated copies of the two universes  spliced together cyclically along the entanglement cut, here the entirety of universe $A$.   Normally the $n$ replicas of universe $B$ would simply go along for the ride.  But now we allow for  Euclidean wormhole solutions, connecting some of the copies of the gravitating universe.   Including these wormholes, we find that the entanglement entropy of universe $A$ minimizes a functional that includes the surface area of a disconnected island in the gravitating universe $B$.\footnote{In two dimensional JT gravity the ``surface area'' of an interval is determined by the value of the dilaton at the endpoints.}

In Sec.~\ref{eq:backreaction} we illustrate these results  in an explicit example with a negative cosmological constant in universe $B$.  Tracing over universe $A$ leaves the field theory in universe $B$ in a mixed state. The stress tensor of this state backreacts on the geometry in $B$ following JT gravity to lengthen the throat behind the black hole horizon, thereby creating a causal shadow \cite{Bak:2018txn}.\footnote{In detail, there is a long Lorentzian wormhole connecting two horizons, and the inner part of the wormhole lies in a causal shadow, namely, it is disconnected from both asymptotic boundaries of $B$.}   The causal shadow behind the horizon contains the island in $B$ that appears in the entropy functional for $A$.   This island increases in size as the entanglement between $A$ and $B$ grows, until it eventually fills up the causal shadow and hence the black hole interior on the $t=0$ Cauchy slice. At that point its boundary is the black hole horizon.  In this high temperature regime, the surface area of the island computes the entropy of the black hole in $B$, and the entanglement entropy of $A$ is determined by the minimum of this quantity and the naive entropy of the field theory quanta in $A$, leading to an analog of the Page transition for evaporating black holes.

We conclude in Sec.~\ref{eq:discussion} with an argument that classical correlation between two universes can not create an island on the gravitating universe.  We also provide  a holographic construction of our entropy formula, in which the field theory on each of the  universes $A$ and $B$ is taken to be conformal with large central charge.  In this case, the state on the bipartite system  is precisely the thermofield double, and the gravitating geometry on $B$ is global AdS$_2$. We now suppose that the entangled CFTs have a dual 3d gravitational description in terms of an eternal BTZ black hole whose boundaries coincide with the original 2d universes. (We want these boundaries to be segments, not circles, so we also introduce end-of-the world branes to truncate the geometries.) In this setting, the CFT entanglement entropies can be computed via the Ryu-Takayanagi formula as the lengths of geodesics in the BTZ bulk. We find that the bulk entanglement wedge of the boundary non-gravitating universe is defined by geodesics that terminate at the two endpoints of the  causal shadow region in the gravitating universe $B$.  In the high temperature limit, the  growth of the eternal black hole in $B$ along with its interior causal shadow region  then implies that the  bulk geodesics defining the entanglement wedge are pushed to the boundary. Thus the entire universe $B$ lies in the entanglement wedge of $A$ and can be reconstructed from the data in $A$.

Previous results on the appearance of gravitating islands in the entropy functional have focused  on black holes, and  sometimes involve explicit models of microstates \cite{Penington:2019kki,Balasubramanian:2020hfs}.  But, as in \cite{Almheiri:2019qdq}, we do not need  microstates to see the appearance of entanglement islands.  The key ingredient is actually monogamy of entanglement.  In our situation, we have field theories on universes $A$ and $B$, with quantum gravitational degrees of freedom (which we do not model explicitly) in  $B$.  These three systems are entangled as follows: the gravitational degrees of freedom and field theory on $B$ are entangled due to their presence in the same spacetime, and the field theory degrees of freedom on $A$ and $B$ are entangled by our choice of state. Thus, as we tune the entanglement between the two field theories by adjusting the state, we expect monogamy to decrease the entanglement between the degrees of freedom on $B$.  This will lead to large effects on the structure of the entanglement entropy of $A$.  The striking lesson is that  semiclassical gravity has direct access to the large corrections which result from monogamy without any explicit reference of the microstates of quantum gravity.  Our construction can be used to study this effect cleanly in cosmological settings \cite{WI}.\footnote{See \cite{Dong:2020uxp,Chen:2020tes,Hartman:2020khs,VanRaamsdonk:2020tlr} for recent work on cosmological islands.  Also see \cite{Chen:2019uhq,Chen:2020uac,Chen:2020jvn,Rozali:2019day,Sully:2020pza,Liu:2020gnp,Liu:2020jsv,Hollowood:2020cou,Hollowood:2020kvk,Banks:2020zrt,Geng:2020qvw,Krishnan:2020fer} and the review \cite{Almheiri:2020cfm} (and references therein) for more general discussions of the island formula in various contexts.  An outgrowth of these developments is a resurgence of interest in baby universes, which emerge naturally from Euclidean wormholes and appear to imply an ensemble interpretation of gravity \cite{Saad:2019lba,Marolf:2020xie} (see \cite{Balasubramanian:2020jhl} for more comprehensive references and a discussion of the main ideas in a simple model).}

\section{Entanglement between two universes
and complementary islands}  \label{sec:setup}

We consider two disconnected universes,  $A$ and  $B$, both with Cauchy surfaces and both carrying quantum field theories. These two theories can be different in general, but for simplicity we will assume that they are two identical conformal field theories. In addition, we turn on semiclassical gravity on $B$, but not $A$.  Therefore, the schematic effective actions on $A$ and $B$ are
\begin{equation}
    \log Z_A = \log Z_{\text{CFT}}[A] , \hspace{.5cm} \log Z_B = -I_\text{grav}[B] + \log Z_{\text{CFT}}[B] .
\end{equation}
When we search for combined saddle-points of the gravity-plus-field theory system, we will solve the gravitational equations of motion in $B$ with a source given by the quantum expectation value of the stress tensor in some state that is entangled with the field theory $A$.

The total Hilbert space of the field theories is a tensor product $\mathcal{H}_A \otimes \mathcal{H}_B$.  The two disconnected universes  cannot communicate classically, but may be entangled quantum mechanically.  We will consider the  entangled state
\be 
|\Psi \ra = \sum^{\infty}_{i=1} \s{p_{i}} | i\ra_{A}\; \otimes  | \psi_{i} \ra_{B},  \qquad p_{i} =\f{e^{-\beta E_{i}}}{Z(\beta)}. \label{eq:HHstate}
\ee
The states $\ket{i}_A$ are  energy eigenstates of CFT$_{A}$.
To define $\ket{\psi_i}_B$, we imagine placing the CFT$_{B}$ in an energy eigenstate with energy  $E_{i} $. This energy may include a contribution from the conformal anomaly if the background geometry is curved.    So the state resembles a thermofield double, but the weighting factors in the sum depend on the energy of the component in the gravitating universe $B$.\footnote{This setup can be generalized to have different field theories on $A$ and $B$.  For example, we can order the  eigenstates of the field theory Hamiltonians on $A$ and $B$ by their energy eigenvalues, and then entangle them so that the product of the i$^{\rm th}$ states in this ordering appear in the sum \eqref{eq:HHstate}.}
We can solve the gravitational equations of motion using the CFT stress tensor expectation value in this state.  Therefore, we think of  $\ket{\psi_i}_B$ as a quantum CFT state in the corresponding back-reacted geometry.\footnote{Note that this differs from previous work on similar states \cite{Penington:2019kki,Balasubramanian:2020hfs}, as we are not defining $\{ \ket{\psi_i}_B \}$ to be a set of black hole microstates drawn at random from some microcanonical energy window.}
The states $\ket{\psi_i}_B$ form a basis for the effective field theory on $B$, but need not model the black hole microstates. $Z(\beta) = \sum_{i} e^{-\beta E_{i}}$ is the CFT partition function on $B$.

\begin{figure}[t]
    \centering
    \includegraphics[scale=.25]{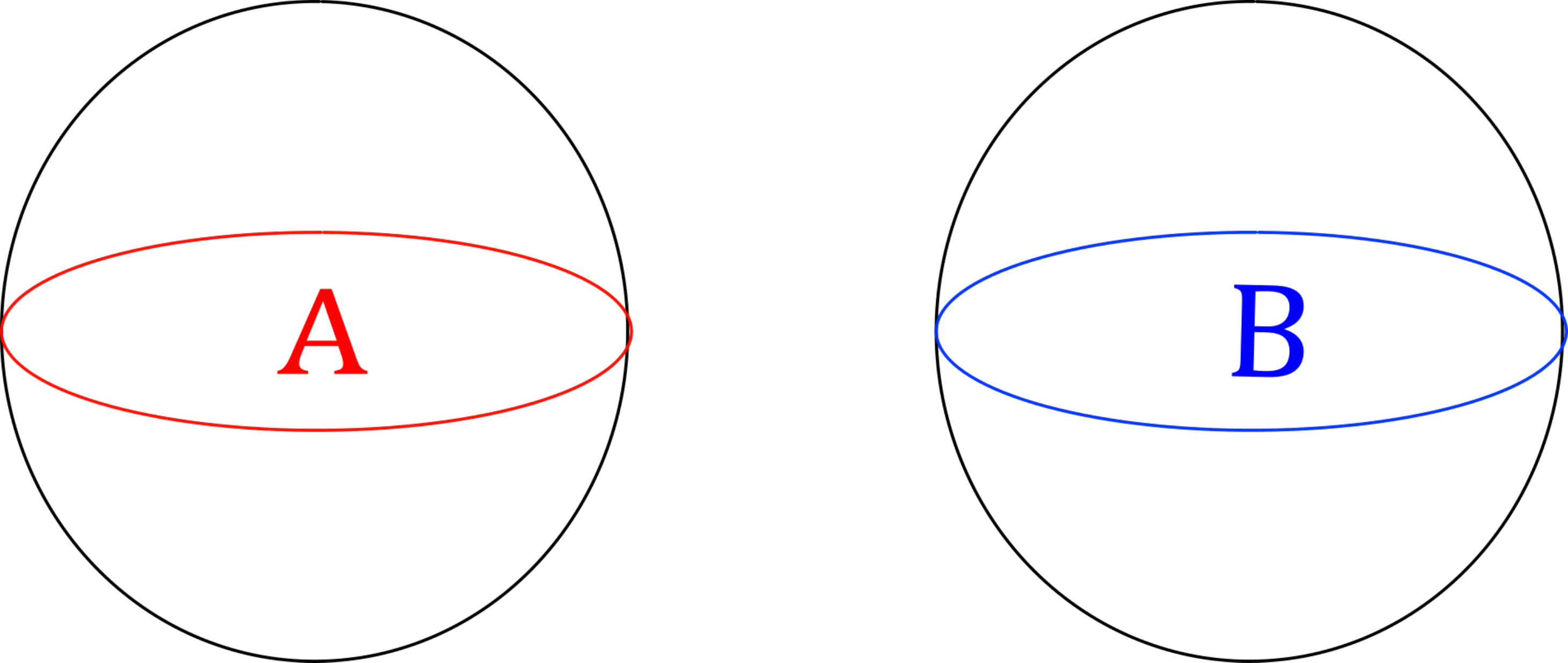}
    \caption{\small{We consider the entanglement between two disjoint universes, $A$ and $B$.  Both universes have propagating matter, but only $B$ gravitates.   The universes may be closed and bounded as illustrated, or non-compact.}}
    \label{fig:Setup}
\end{figure}

The state (\ref{eq:HHstate}) has a one-parameter dependence on an inverse temperature $\beta$, which can be tuned to change the strength of entanglement between $A$ and $B$.    Since the CFT degrees of freedom on $B$ must also be entangled with gravitational degrees of freedom, we can probe the unitarity of quantum gravity by forcing the entanglement between CFT fields on $A$ and $B$ to be large. By doing this, we will see that monogamy of entanglement between the gravitational degrees of freedom on $B$, matter fields on $B$, and  matter fields on $A$ will  lead to large modifications of the entanglement entropy of the field theory state \eqref{eq:HHstate} despite the fact that the gravitational microstates are not explicitly included.

\subsection{A general argument}

The structure of the ``island formula'' in \cite{Almheiri:2019hni,Almheiri:2019qdq}, suggests that there should be two phases for the entanglement entropy $S(A)$.  When the entanglement is sufficiently weak, the entropy should be equal to the  von Neumann entropy of the thermal density matrix on $A$, $S_{\beta}(A)$.  Because we have a pure state on $A$ and $B$ together, it follows that  $S_{\beta}(A) = S_{\beta}(B)$ (since the reduced state on $B$ is also thermal).   When the entanglement is large, $S(A)$ should have contributions from the surface area of an ``island'' $C \subset B$  ($\text{Area}(\partial C)/4G_N$) and from the entropy of the field theory on $A \cup C$ ($S_{\beta}(A\cup C)$).  Since the total state on $AB$ is pure, we can equally well compute this island phase entropy of region $AC$ by computing the thermal entropy of the complement of the island $\overline{C} \subset B$, along with the same endpoint area contributions (since $\partial C = \partial \overline{C}$). All told, the island phase should yield an expression like $\frac{\text{Area}(\partial \overline{C})}{4G_N} + S_{\beta}(\overline{C})$.   

The transition between the no-island and island phases is determined by asking which entropy expression gives the smaller value.   Unlike in \cite{Almheiri:2019hni,Almheiri:2019qdq}, the entropy $S_{\beta}(A)$ is manifestly UV-finite, since  we are considering  entropy on a complete Cauchy slice (so that there are no endpoint UV divergences), on which states have been excited with some finite effective temperature.  Thus, in order for the island phase to dominate, the correct expression for the entanglement entropy in this regime cannot simply involve $S_{\beta}(A\cup C) = S_{\beta}(\overline{C})$ because this quantity will have the standard endpoint divergences in a quantum field theory. Since these divergences arises from  UV modes that straddle a cut, a natural way to remove them is  to subtract the entanglement entropy in the vacuum:
\begin{equation}
    S(A) = \min 
    \begin{cases}
    S_{\beta}(B) ,&\\
    \underset{C}{\min} \left[ \frac{\text{Area}(\partial \overline{C})}{4G_N} + S_\beta(\overline{C}) - S_{{\rm vac}}(\overline{C}) \right] .
    \end{cases}
\end{equation}
We will see that precisely this expression results from applying the replica method to JT gravity coupled to a 2d CFT.

\subsection{Evaluation of the R\'enyi entropy}

We are interested in the entanglement entropy of the reduced density matrix $\rho_A$, which is obtained by tracing out $\mathcal{H}_B$ from the total state \eqref{eq:HHstate}.  
Since the states $\ket{i}_A$ are orthonormal, the R\'enyi entropies are computed by
\be 
{\rm tr} \; \rho_{A}^{n} = \sum_{i_{1} \cdots i_{n}} p_{i_{1} } \cdots p_{i_{n} } \la \psi_{i_{1}} | \psi_{i_{2}} \ra \la \psi_{i_{2}} | \psi_{i_{3}} \ra  \cdots \la \psi_{i_{n}} | \psi_{i_{1}} \ra. \label{eq:renyin}
\ee
Note that the states $|\psi_i\rangle$ should be regarded as quantum CFT states in a backreacted geometry, and so the overlaps on the right hand side of \eqref{eq:renyin} will include  contributions from both the field theory and from the semiclassical gravity.  Below, we focus on the cases where the universe $B$ is either a sphere (two-dimensional Euclidean de Sitter) or a disk (Euclidean 
anti de Sitter).  The discussion  is  parallel for these two cases,  because a disk is just half of  a sphere, and so we will focus mostly on the sphere and then discuss the disk in analogy.  Universe $B$ will contain a 2d CFT coupled to Jackiw-Teitelboim (JT) gravity described by the action
\begin{equation}
    \log Z_B 
    = \frac{\phi_0}{4\pi} \left[ \int_B R + \int_{\partial B} 2K \right] + \int_B \frac{\Phi}{4\pi} (R-\Lambda) + \frac{\Phi_b}{4\pi} \int_{\partial B} 2K +
    \log Z_{\text{CFT}} [g],
\label{eq:ZB}
\end{equation}
where $g$ is the metric on $B$, $\Phi_b$ is the boundary value of the dilaton $\Phi$, $\Lambda$ is the cosmological constant, and $\phi_0$ is the ground state entropy of a higher-dimensional black hole whose dimensional reduction leads to the JT theory.
We have set $4G_N = 1$.

To compute the overlap $ \la \psi_{i} | \psi_{j} \ra $ between two states on the sphere, we first prepare the excited state $|\psi_{j} \ra$ on the equator  by a path integral on the southern  
hemisphere  with an insertion of a local operator $\psi_j (0)$ at the south pole. We prepare $ \la \psi_{i}|$ similarly by insertion of  $\psi_i (\infty)$ at the north pole.  So the overlap between these states equals the two point function on the sphere:
$ \la \psi_{i} | \psi_{j} \ra = \la  \psi_{i} (\infty) \psi_{i} (0) \ra$. This implies that the Renyi entropy \eqref{eq:renyin} has a path integral  representation on  $n$ copies of the gravitating universe. Naively each copy is totally disconnected from the others.  But the standard rules of the Euclidean gravitational path integral state that we must sum over all topologies that are consistent with the boundary conditions.  If these rules also apply to gravitational replicas, then we must also consider the contribution of replica manifolds connected by euclidean wormholes.  In the semiclassical limit, the path integral will then be dominated by sum over  replica-symmetric saddleppoints in each of these sectors.

The calculation proceeds in three steps. We start from a fixed gravitational configuration, specified by a profile $( \phi, g_{\mu\nu})$, which is in general off shell. In principle, there are an enormous number of topologies satisfying the desired boundary conditions, with wormholes connecting some replicas and not others.
We will focus on two specific replica-symmetric saddles,
namely the fully disconnected saddle, $M_{\text{disc}}$ where each copy is fully disconnected from the others, and the fully connected one, $M_{\text{conn}}$ where all copies are connected to others by Euclidean wormholes. Later we will see that $M_{\text{disc}}$  gives the dominant contribution  in the low entanglement temperature limit $\beta \rightarrow \infty$. 
Similarly $M_{\text{conn}}$ dominates the answer in the high temperature limit (Fig.~\ref{fig:sphere-replicas}). Next we compute the required CFT correlators on the fixed gravitational background.  This task is particularly simple in JT gravity because, from the action \eqref{eq:ZB}, they depend only on the metric $g_{\mu\nu}$ and {\it not} on the dilaton profile $\phi$. In addition, in JT gravity the equation of motion derived from \eqref{eq:ZB} implies that only the  dilaton is affected by the back reaction of the stress energy tensor.  This implies that  CFT correlation functions can be evaluated without including  back reaction,  and are evaluated in the fixed curvature metric implied by the JT gravity equations of motion.  Finally, we compute the dilaton  profile which accounts for the back reaction from the CFT stress tensor, and then evaluate the on shell action.  Putting everything together gives
\be
{\rm tr} \rho^{n}_{A} =\f{1}{Z_{1}^{n}}\left( e^{-I_{\text{grav}} [M_{\text{disc}}]} Z_{{\rm CFT}} [M_{\text{disc}}]+e^{-I_{\text{grav}} [M_{\text{conn}}]} Z_{{\rm CFT}} [M_{\text{conn}}]\right) ,
\label{eq:renyiform1}
\ee
where $Z_{1}$ is the normalization,
\be 
Z_{1}= e^{-I_{\text{grav}}[M_{1}]} Z(\beta)  .
\ee
Here $M_1$ is the saddlepoint geometry for a single copy of the gravitating universe and comes from our normalization of the gravitational part of the path integral to ensure that $\tr(\rho_A) = 1$. In \eqref{eq:renyiform1} the $Z_{\text{CFT}}$ factors are a schematic notation for the weighted sums of CFT correlation functions required to compute the overlaps in \eqref{eq:renyin}, evaluated on the disconnected and connected saddles respectively. These sums will have an overall factor of $1/Z(\beta)^n$ from the normalization of the $p_i$, which we have pulled out and included in $Z_1$.

When universe $B$ has the disk topology, the n-fold replica has n disconnected circles as its boundary.  The disconnected and connected saddles that compete are then $n$ disconected disks versus a single multi-boundary replica wormhole.  In this case, the connected wormhole conributions will be topologically suppressed by the exponential of the Euler character.  When the universe $B$ has the  topology of the sphere, the situation is more subtle.  The sphere is compact, so there is no replicated boundary condition per se.   The natural candidate for the connected saddlepoint wormhole saddlepoint in this case is the branched sphere.\footnote{See \cite{Dong:2020uxp} for a similar situation.}   While the branched sphere has the same topology as the sphere itself, it has the the cyclic  symmetry that would be required for it to contribute as a replica saddlepoint.  But because it is topologically the same as the sphere, it will have a ``geometric'' suppression, rather than a topological suppression, as we will see shortly.

\subsection{Disconnected saddlepoint} 
We first evaluate the disconnected saddlepoint $M_{\text{disc}}$, where the replica copies of universe $B$ are disjoint.  The overlaps $\inner{\psi_i}{\psi_j}$  in \eqref{eq:renyin} are then computed by evaluating the path-integrals on these disconnected factors.  The gravitational saddlepoint contribution to the path integral is $I_{\text{grav}} [M_{\text{disc}}]= n I_{\text{grav}}[M_{1}]$.  Meanwhile, the CFT two-point functions in \eqref{eq:renyin}, which are computed on disjoint replica factors, are diagonal ($\inner{\psi_i}{\psi_j} \propto \delta_{ij}$) since the states $|\psi_i\rangle$ are orthornomal energy eigenstates.  Including this diagonal correlator along with the weighting factors $p_i$, the disconnected saddlepoint contribution is
\begin{equation}
\frac{e^{-nI_{\text{grav}}[M_{1}]} \sum_i e^{-n\beta E_i}}{Z_1^{n}} \, ,
\end{equation}
where we included the normalization of the gravitational path integral from \eqref{eq:renyiform1}.  Thus the R\'{e}nyi entropies collapse to
\begin{equation}
    \frac{Z(n\beta)}{Z^n(\beta)} .
    \label{eq:QFTresult}
\end{equation}

\subsection{Connected saddlepoint}

We next evaluate the ``fully connected" saddle $M_{\text{conn}}$.
To construct $M_{\text{conn}}$, we introduce a segment $C_x$ on a Cauchy slice of $B$, with linear size $2\pi x$. 
We then cyclically glue the $n$ disconnected components of $M_{\text{disc}}$ along $C_x$, forming a $\mathbb{Z}_n$-symmetric Euclidean wormhole topology which obeys the gravitational equations of motion.  This construction is similar to the bulk replica manifold which appears in \cite{Lewkowycz:2013nqa}, though there are $2n$ CFT operators in various positions on our manifold, and we are not assuming the existence of a boundary. Fig.~\ref{fig:sphere-replicas} shows
an $n=3$ example of $M_{\text{conn}}$ with $B$ a 2-sphere.

\subsubsection{Gravitational contribution}
To compute the contribution to the R\'{e}nyi entropy from the connected saddlepoint we have to calculate the gravitational action on $M_{\text{conn}}$, which is an n-fold cover of the sphere.  However, to compute the entanglement entropy we have to be able to analytically continue in $n$.  A good prescription for doing this was given in \cite{Lewkowycz:2013nqa}.  Following, \cite{Lewkowycz:2013nqa} we work with the quotient of this space by the replica symmetry $Q_{n}=M_{\text{conn}}/Z_{n}$.    The $Z_n$ replica symmetry has two fixed points in its action on $Q_{n}$ leading to conical singularities with the opening angle $2\pi/n$. This implies the quotient does not solve the gravitational equations of motion at the conical singularities. Nevertheless the conical space is useful because there is a simple relation
\be 
I_{{\rm grav}}[M_{\text{conn}}]=n I_{{\rm grav}}[Q_{n}] ,
\label{eq:replicaaction}
\ee 
due to the locality of the action, where
on the right hand side we evaluate the action {\it without} the contribution of the conical singularity.  The quotient manifold $Q_n$ can be analytically continued to any $n$ because it is just a sphere with two conical defects, which can have any real opening angle determined by a parameter $n$.   Let $(\theta, \phi )$ be the coordinates on this sphere $Q_{n}$ with the metric
\be 
 ds^{2} =d\theta^{2} +\sin^{2} \theta d \phi^{2}. \label{eq:sphere}
\ee
The conical singularities of $Q_n$ are located at $(\theta,\phi) = (\frac{\pi}{2},0)$ and $(\theta,\phi) = (\frac{\pi}{2},2\pi x)$, with $0<x<1$.

In practical terms, the computation of the right hand side of \eqref{eq:replicaaction} can be implemented in two ways: (1) we can prepare a small circle which surrounds the conical singularity and impose a boundary condition on it; (2)  we can introduce a cosmic string which gives a source for the conical defect and renders it on-shell  \cite{Almheiri:2019qdq}.  In either case, if we evaluate the total action the conical singularity does not contribute to the action at all. Both methods give the same result because, (1) the CFT part of the path integral is only sensitive to the conformal class of  the metric, and not to the boundary condition or the cosmic brane, and (2) the
 gravitational contribution in both cases is identical.  Following either prescription in JT gravity gives  $I_{{\rm grav}} = (n-1) \left(\Phi(\frac{\pi}{2},0) +\Phi(\frac{\pi}{2},2\pi x) \right)$ \cite{Almheiri:2019qdq}.

\subsubsection{CFT contribution}

\begin{figure}
     \centering
     \includegraphics[scale=.2]{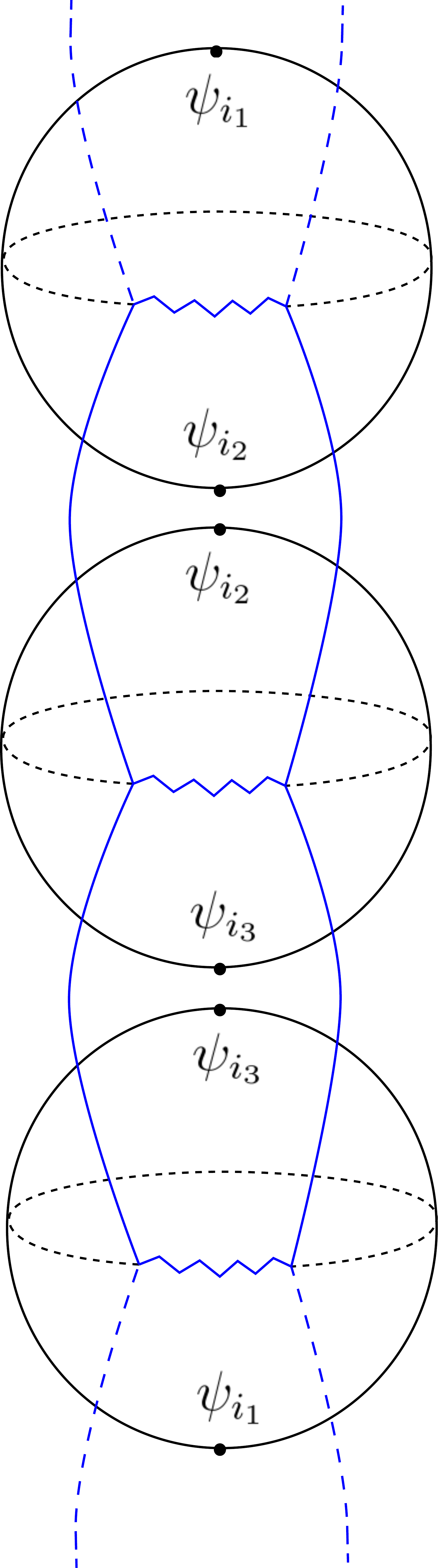}
     \hspace{3cm}\includegraphics[scale=.2]{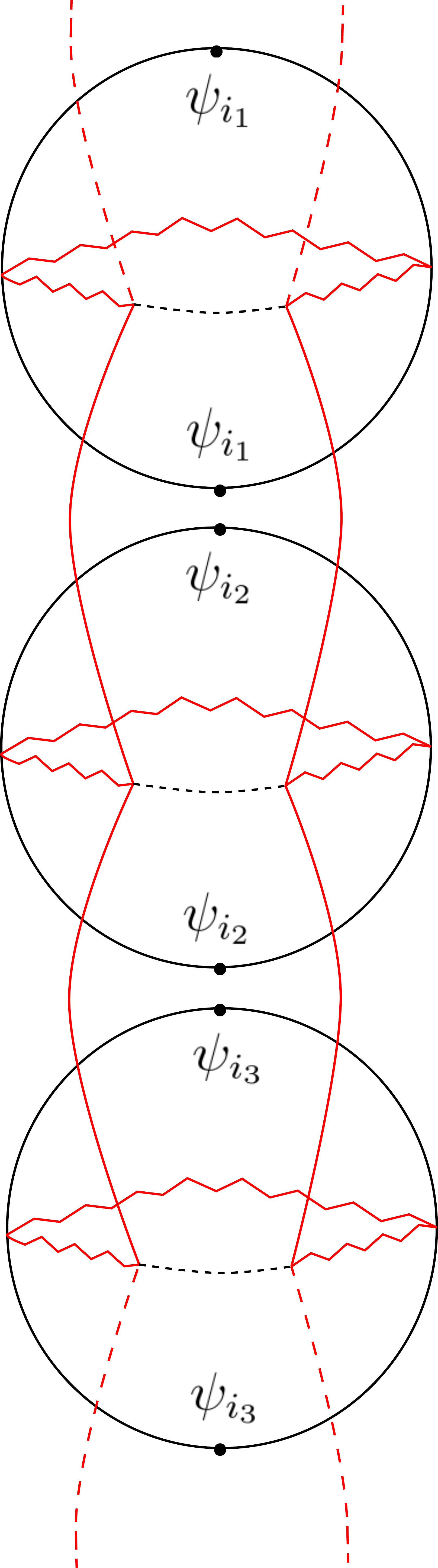}
     \caption{\small{Left: The replica manifold $M_{\text{conn}}$ for $n=3$, with operator insertions which differ on the same sheet.  The blue wiggly line is the cut $C_x$ with length $2\pi x$, and the spheres are glued cyclically along it.
     Right: the replica manifold $M_{\text{conn}}$, with operator insertions that are the same on a given sheet.  The spheres are again glued cyclically, this time along the red wiggly line $\overline{C_x}$, which is a cut of length $2\pi(1-x)$.
     The path integrals on these two manifolds with the specified operator insertions are equivalent, as proven in Appendix~\ref{sec:identity}.
     }}
     \label{fig:sphere-replicas}
 \end{figure}

The product of  overlaps that appears in the R\'{e}nyi entropy \eqref{eq:renyin} can be written in terms of a 2n-point correlation function on $M_{\text{conn}}$:
\be 
\la \psi_{i_{1}} | \psi_{i_{2}} \ra \la \psi_{i_{2}} | \psi_{i_{3}} \ra  \cdots \la \psi_{i_{n}} | \psi_{i_{1}} \ra=\avg{\psi_{i_1}(\infty_1) \psi_{i_2}(0_1) \dots \psi_{i_n}(\infty_n) \psi_{i_1}(0_n)}_{M_{\text{conn}}}.
\label{eq:corrsig}
\ee
The metric the smooth manifold $M_{\text{conn}}$ and the n-fold branched cover of the sphere $\Sigma_{n}$ are related by a Weyl transformation, and one can choose the Weyl factor $\Omega =1$ except in a small neighborhood of the branch points \cite{Faulkner:2013ana}.  We can take  $\Sigma_n$ to have the standard sphere metric  \eqref{eq:sphere}. Then the cut along which branches are sewn is located at $C_{x}: \theta=\f{\pi}{2},\; 0<\phi < 2\pi x$, with $0<x<1$. By the state-operator correspondence the overlaps  in \eqref{eq:corrsig} can be computed by placing the appropriate operators at the north and south poles of the various spheres in $M_{\text{conn}}$.  These poles can be chosen so that they are far from the branch points after mapping to $\Sigma_n$.   This means that the correlation function on the right hand side of \eqref{eq:corrsig} is equal to the one on $\Sigma_{n}(C_{x})$,
\be 
\avg{\psi_{i_1}(\infty_1) \psi_{i_2}(0_1) \dots \psi_{i_n}(\infty_n) \psi_{i_1}(0_n)}_{M_{\text{conn}}}=
\avg{\psi_{i_1}(\infty_1) \psi_{i_2}(0_1) \dots \psi_{i_n}(\infty_n) \psi_{i_1}(0_n)}_{\Sigma_{n}(C_x)} \, ,
\label{eq:correeq}
\ee 
where by $0_k$ and $\infty_k$  in the right hand side we label the operator insertions on the north and south poles of the $k^{\text{th}}$ sheet of the branched sphere $\Sigma_{n}(C_{x})$ (see Fig.~\ref{fig:sphere-replicas}).  (Note that in this notation we are mapping the sphere to the plane so that the north and south poles map to infinity and the origin respectively.)

The correlation functions $\la \cdots \ra_{\Sigma_{n}(C_x)}$ on the branched sphere have been studied in  calculations of  the entanglement entropy  of excited states, e.g., in \cite{Asplund:2014coa,Sarosi:2016oks}.
In deriving the identity \eqref{eq:correeq}, one might worry about the Weyl anomaly when one of the local operators is the stress tensor,  $\psi_{i} =T$.  However, because we are choosing the the Weyl factor $\Omega =1$ except in a small neighborhood of the branch points, insertions at the poles will not be affected.

The properties of the  correlation functions $\la \cdots \ra_{\Sigma_{n}(C_x)}$  as a function of the wormhole size $x$ can be studied by mapping $\Sigma_{n}(C_x)$ to a unformized plane.  When the size of the wormhole is small $x \ll 1$,  $\psi_{i_{k}} (\infty_{k}) \rightarrow \psi_{i_{k+1}} (0_{k})$ on the uniformized plane \cite{Sarosi:2016oks}, the correlation function factorizes into a  product of overlaps, 
 \be 
 \avg{\psi_{i_1}(\infty_1) \psi_{i_2}(0_1) \dots \psi_{i_n}(\infty_n) \psi_{i_1}(0_n)}_{\Sigma_{n}(C_x)} \rightarrow  \la \psi_{i_{1}} | \psi_{i_{2}} \ra \la \psi_{i_{2}} | \psi_{i_{3}} \ra  \cdots \la \psi_{i_{n}} | \psi_{i_{1}} \ra, \quad x \rightarrow 0.
 \ee 
 In this limit, the CFT contrbution is thus the same as from the disconnected saddlepoint and so is given by the ratio of thermal partition functions.  This gives a contribution to the R\'{e}nyi entropy of the form
 \be 
 \frac{1}{Z_1^n} e^{-I_{\text{grav}} [M_{\text{conn}}]} Z_{{\rm CFT}}[M_{\text{conn}}] \rightarrow  \left( e^{-I_{\text{grav}} [M_{\text{conn}}] + nI_{\text{grav}}[M_1]} \right) \f{Z(n\beta)}{Z^{n}(\beta)} .
 \ee
On the other hand, when the wormhole is large, $x \rightarrow 1$  a different OPE channel dominates ($\psi_{i_{k}} (\infty_{k}) \rightarrow \psi_{i_{k}} (0_{k})$) \cite{Sarosi:2016oks}, so that 
\be 
\avg{\psi_{i_1}(\infty_1) \psi_{i_2}(0_1) \dots \psi_{i_n}(\infty_n) \psi_{i_1}(0_n)}_{\Sigma_{n}(C_x)} \rightarrow  \la \psi_{i_{1}} | \psi_{i_{1}} \ra \la \psi_{i_{2}} | \psi_{i_{2}} \ra  \cdots \la  \psi_{i_{n}} | \psi_{i_{n}} \ra=1 .
\label{eq:correq1}
\ee
Thus, in this limit the CFT contribution $Z_{{\rm CFT}}[M_{\text{conn}}]$
becomes independent of $\beta$ because the sum of the correlation functions \eqref{eq:correq1} weighted by $p_i$ as in \eqref{eq:renyin} equals 1.   We will see that this will lead a saturation of entanglement entropy at high entanglement temperatures ($\beta \to 0$).

In fact we can compute the result for any size of the wormhole $x$. To to do this we can use the identity in Appendix \ref{sec:identity},
\begin{equation}
    \avg{\psi_{i_1}(\infty_1) \psi_{i_2}(0_1) \dots \psi_{i_n}(\infty_n) \psi_{i_1}(0_n)}_{\Sigma_{n}(C_x)} = \avg{\psi_{i_1}(\infty_1) \psi_{i_1}(0_1) \dots \psi_{i_n}(\infty_n) \psi_{i_n}(0_n)}_{\Sigma_{n}(\overline{C_x})} \label{eq:corridentity},
\end{equation}
where on the right hand side operators of the same type are located on the same sheet, and the cut has been moved to the complement subregion $\overline{C_x}$, so $C_x \cup \overline{C_x}$ is the total Cauchy surface.  Intuitively, we can visualize the action of the  conformal transformation which produces this relation  by pulling the cut $\overline{C_x}$ through the insertion at the origin. If $B$ is a 2-sphere as in Fig.~\ref{fig:sphere-replicas}, the region $C_x$ is an interval on the equator of size $2\pi x$ and the complement region $\overline{C_x}$ is the complementary interval of size $2\pi(1-x)$.  Making use of this identity, we can perform the sums over operators in \eqref{eq:renyin} for the connected contribution to the R\'{e}nyi entropy.
Before performing the resummations, the right hand side of \eqref{eq:corridentity} becomes
\begin{equation}
    \avg{\psi_{i_1}(\infty_1) \psi_{i_1}(0_1) \dots \psi_{i_n}(\infty_n) \psi_{i_n}(0_n)}_{\Sigma_{n}(\overline{C_x})} = \frac{\tr [ \rho_{\psi_{i_1},\overline{C_x}} \dots \rho_{\psi_{i_n},\overline{C_x}} ]}{\tr \rho^n_{\text{vac},\overline{C_x}} } ,
\end{equation}
where the $\rho_{\psi_i,\overline{C}}$ is the reduced density matrix on $\overline{C}$ of the excited CFT state $\ket{\psi_i}\bra{\psi_i}$.
Then, the sums over the indices $i_1, \dots , i_n$ in \eqref{eq:renyin} together with the weighting factors $p_i$ will resum these excited state density matrices into thermal density matrices.
All together, the trace of the product of $n$ excited state density matrices turns into the $n^{\text{th}}$ R\'{e}nyi entropy of a thermal density matrix:
\begin{equation}
     \frac{1}{Z_1^n} e^{-I_{\text{grav}} [M_{\text{conn}}]} Z_{{\rm CFT}}[M_{\text{conn}}] \to \left( e^{-I_{\text{grav}}[M_{\text{conn}}] + nI_{\text{grav}}[M_1]} \right) \frac{\tr \rho^n_{\beta,\overline{C_x}}}{\tr\rho^n_{{\rm vac},\overline{C_x}}}  ,
    \label{eq:gravrenyi}
\end{equation}
where the Boltzmann factors in \eqref{eq:renyin} have combined with the matching operator insertions on a given sheet to form the R\'{e}nyi entropy of the region $\overline{C_x}$ in the thermal density matrix.
Note that \eqref{eq:gravrenyi} holds for all $x$, not just in a specific OPE limit.
Here the reduced thermal CFT density matrix at inverse temperature $\beta$ on subregion $X$ is denoted $\rho_{\beta,X}$, and the vacuum density matrix is obtained when $\beta = \infty$.
The CFT contribution then becomes a ratio of this thermal subregion R\'{e}nyi entropy and a zero temperature normalization factor $\rho_\infty = \rho_\text{vac}$. The final result for the R\'{e}nyi entropy $\tr \rho_A^n$ due to our replica-symmetric saddle contributions is
\begin{equation}
   \tr \rho_A^n = 
     \frac{Z(n\beta)}{Z^n(\beta)} +
    e^{-I_{\text{grav}}[M_{\text{conn}}]} \frac{\tr \rho^n_{\beta,\overline{C_x}}}{\tr\rho^n_{{\rm vac},\;\overline{C_x}}} .
    \label{eq:totalrenyi}
\end{equation}

We know exact expressions for the R\'enyi entropy of a thermal ensemble in a  2d CFT in the large $c$ limit,\footnote{We use the holographic expressions \cite{Hubeny:2007xt} for thermal R\'enyi entropies, though we expect that our result to be be unchanged if we pick a different sort of CFT (analogous to \cite{Almheiri:2019qdq}, where a free fermion theory was used). The emergence of islands should be essentially universal, as long as we have a parameter with which to increase entanglement without bound (in our case, the CFT temperature).}
\be 
 \tr \rho^n_{\beta,\overline{C_x}}=\left( \sinh \f{\pi L  (1-x)}{\beta}\right)^{-\Delta_{n}}, \quad \tr \rho^n_{{\rm vac},\;\overline{C_x}} =\left( \sin  \f{\pi^{2} x}{L} \right)^{-\Delta_{n}} , \quad \Delta_{n} =\f{c}{12} \left(n-\f{1}{n} \right) , \label{eq:renyis}
\ee
where we introduced an explicit dependence on the system size $L$, which is determined so far by the period of the $\theta$, namely $2\pi$.

The entanglement entropy $S_A$ is given in terms of these quantities by
\begin{equation}
    S_A = -\left[ \partial_n  \frac{\log \tr \rho_A^n}{n} \right]_{n=1} .
\end{equation}
The analytic continuation in $n$ of \eqref{eq:totalrenyi}, which is necessary for taking the derivative $\partial_n$ appearing in $S_A$, can be carried out using the usual quotient by $\mathbb{Z}_n$ action \cite{Lewkowycz:2013nqa,Almheiri:2019qdq}.
The result is a double minimization formula, first over the dominant saddle and second over the only modulus in the problem, $x$:
\begin{equation}
    S_A = \min 
    \begin{cases}
    S_{\beta}[B] ,&\\
    \underset{x}{\text{min ext}} \left[ \frac{\text{Area}[\partial \overline{C_x}]}{4G_N} + S_\beta[\overline{C_x}] - S_{{\rm vac}}[\overline{C_x}] \right] .
    \end{cases}
    \label{eq:complement-islands}
\end{equation}
The entanglement entropy of a holographic CFT (which therefore has large $c$) in a thermal state is
\be 
S_{\beta}[\overline{C_x}] =\f{c}{3} \log \left[ \f{\beta}{\pi}\sinh \f{\pi (1-x)}{\beta} \right] , \quad  S_{{\rm vac}} [\overline{C_x}]=\f{c}{3}\log \left[2 \sin \pi x \right] \, .
\ee
The leading order expansion of the expression in $1/\beta$ also gives the universal leading term in the entanglement entropy for any CFT at high temperature.   In addition, in JT gravity
\begin{equation}
    \frac{\text{Area}[\partial \overline{C_x}]}{4G_N} = \phi(\partial \overline{C_x}) ,
\end{equation} 
Putting everything together we can analyze the behavior of $S_A$ as the entanglement temperature changes.

At low entanglement temperature, or large $\beta$, the disconnected gravitational saddle $M_{\text{disc}}$ dominates, and we recover the thermal entropy of fields on $B$.
At High entanglement temperature, or low $\beta$, the connected gravitational saddle $M_{\text{conn}}$ dominates, and we find an analog of the ``island formula" \cite{Almheiri:2019hni}, namely the second line in \eqref{eq:complement-islands} Indeed, since the state \eqref{eq:HHstate} is  pure  on the union of two universes $AB$, the entanglement entropy on $\overline{C_x}$ must satisfy,
\be
S_{\beta}[\overline{C_x}] -S_{{\rm vac}} [\overline{C_x}] =S_{{\rm TFD}}[AC_x] -S_{{\rm vac}} [AC_x],
\ee
where $S_{{\rm TFD}}$ is the entropy computed in the state \eqref{eq:HHstate}, and, similarly ,
\be
\frac{\text{Area}[\partial \overline{C_x}]}{4G_N} = \frac{\text{Area}[\partial AC_x]}{4G_N}.
\ee
So the second line in \eqref{eq:complement-islands} can also be written as
\begin{equation}
    \underset{x}{\text{min ext}} \left[ \frac{\text{Area}[\partial AC_x]}{4G_N} +  S_{{\rm TFD}}[AC_x] -S_{{\rm vac}} [AC_x] \right].
\end{equation}
so that $C_{x}$ in  universe $B$ is the entanglement island in our scenario.

Notice that the bulk thermal entropy contribution in this setting is naturally renormalized by the vacuum entropy contribution $-S_{\rm vac}$ which did not appear in the original island formula.
This had to appear in our discussion since we are computing the entropy of a normalized pure state in a true tensor product Hilbert space. As such, it must be finite, and indeed all of our possible expressions for $S_A$ are manifestly finite.

\subsection{Generalization to the disk}

Our discussions so far  have  assumed that the gravitating universe had the topology of a sphere, and thus is the Euclidean continuation of 2d de Sitter space.  It is easy to generalize our results to the disk, or 2d Euclidean anti-de Sitter space.  To do this, let us treat the sphere which was studied above as two disks glued on a longitude.  We prepare overlaps  in \eqref{eq:renyin} by inserting   local operators at antipodal points of the boundary of the disk at the locations of the north and south pole of the sphere.  The required product of the overlaps then involves $n$ copies of the disk, which may or may not be connected by  Euclidean wormholes.  The contribution of the fully connected wormhole  is given by the correlation function on the branched disk $\Sigma_{n}^{D} (C_{x})$.

An identity similar to \eqref{eq:corridentity} holds for the branched disk,
\begin{equation}
    \avg{\psi_{i_1}(\infty_1) \psi_{i_2}(0_1) \dots \psi_{i_n}(\infty_n) \psi_{i_1}(0_n)}_{\Sigma^{D}_{n}(C_x)} = \avg{\psi_{i_1}(\infty_1) \psi_{i_1}(0_1) \dots \psi_{i_n}(\infty_n) \psi_{i_n}(0_n)}_{\Sigma^{D}_{n}(\overline{C_x})} \label{eq:corridentity-disk},
\end{equation}
because  the branched disk is regarded as  a part of the branched sphere.  Here we have mapped the disk to the upper half plane the north and south poles have been mapped to the origin and infinity.
In the derivation of this identity in Appendix \ref{sec:identity},  we only used a conformal map on the branched sphere, so by the restriction  we can define the map  for the branched disk.   Thus  the CFT part  of the path integral on the fully connected wormhole  is still given by the ratio, 
${\rm tr} \; \rho_{\beta,\overline{C_{x}}}^{n}/{\rm tr} \rho_{{\rm vac},\overline{C_{x}}}^{n}$.   Likewise, the disconnected contribution will still be a ratio of thermal partition sums $\f{Z(n\beta)}{Z^{n}(\beta)}$.

The only difference from the result in the spherical universe is that a time slice of the universe is not a circle, but an interval.  As a result, both ${\rm tr} \; \rho_{\beta,\overline{C_{x}}}^{n}$ and  $\rho_{{\rm vac},\overline{C_{x}}}^{n}$ depend on the boundary condition at the edges of the  interval. However the choice of the boundary conditions only affects these R\'enyi entropies by the factor of the so-called $g$ function which is  state independent  \cite{Azeyanagi:2007qj,Calabrese:2004eu} .   Since this factor is cancelled in the ratio ${\rm tr} \; \rho_{\beta,\overline{C_{x}}}^{n}/{\rm tr} \rho_{{\rm vac},\overline{C_{x}}}^{n}$, the entropy is still given by \eqref{eq:complement-islands}.

The fact that roughly the same formula for the entropy holds in the case of the disk may be surprising.
One main difference is that there are boundary conditions for the interval, which involves the $g$ function referenced above.
This object is independent of the number of intervals, and indeed whether or not the intervals go to the boundary, upon which we compute the reduced density matrix \cite{Azeyanagi:2007qj,Calabrese:2004eu}.
Furthermore, one may wonder why we have not referred at all to twist operators in both this disk case as well as the sphere discussion which appeared previously.
This is because, much like in the calculation of a relative entropy \cite{Sarosi:2016oks}, the derivation involves a trace of density matrices associated with distinct excited states, and there is no replica symmetry in the replicated manifold.
Therefore, twist operators are not convenient, and we opt instead to perform the path integral with operator insertions by passing to the uniformized plane, at which point we find a $2n$-point correlation function of CFT fields.
Finally, the presence of the boundary may lead one to believe that there ought to be images associated with the operator insertions.
However, since the operator insertions which create the excited states live on the boundary, there are no associated images.
And, since twist operators have not appeared at all in the calculation, there are no images of these either.

\section{Finding the island} 
\label{eq:backreaction}

In JT gravity with a negative cosmological constant, the equation of motion obtained by variation of the dilaton reduces to
\begin{equation}
    R + 2 = 0 ,
    \label{eq:JT-EOM-metric}
\end{equation}
which fixes the curvature to be constant and negative.
This  is true even in the presence of extra matter fields which can backreact on the geometry, since the matter action is independent of the dilaton by assumption. 
For the Euclidean disk topology, this means that any classical solution can be expressed as a portion of the hyperbolic disk.
Therefore, semiclassical solutions of JT gravity on the disk are specified by a matter stress tensor expectation value $\avg{T_{\mu\nu}}$ and a dilaton profile $\phi$.\footnote{In the quantum description of JT gravity, the dilaton profile is often exchanged for a cutoff profile $t(u)$ which is sometimes called the reparametrization (or Schwarzian, due to its effective action) mode \cite{Maldacena:2016upp}.}

To find the dilaton profile  which appears in the semiclassical entropy formula  \eqref{eq:complement-islands}, we must account for
the backreaction of the CFT stress energy tensor by solving the equations of motion of JT gravity coupled to a quantum CFT.
Importantly, only the dilaton field will feel the effects of backreaction, as the metric is completely fixed by \eqref{eq:JT-EOM-metric}.

We consider a CFT on strip, and need to find the backreaction of the CFT stress energy tensor on the dilaton.  
In this section we will work in Lorenzian signature. In doing so, it  is convenient to work in conformal gauge where the metric of the strip is given by \cite{Bak:2018txn}
\be 
ds^{2} =-e^{2\omega} dx^{+} dx^{-} , \quad e^{2\omega} =\f{1}{\cos^{2} \mu}, \quad x^{\pm}=t \pm \mu , \quad -\f{\pi}{2} \leq \mu \leq \f{\pi}{2}. \label{eq:metric}
\ee
The dilaton  satisfies the equations of motion \cite{Almheiri:2014cka}
\begin{equation} 
\begin{split}
2\p_{+} \p_{-} \Phi +e^{2\omega} \; \Phi &=16\pi G \;\la \Psi|T_{+-}|\Psi \ra,  \\
 e^{2\omega} \p_{+} \left[  e^{-2\omega} \p_{+} \Phi \right] &=-8\pi G \; \la \Psi|T_{++} |\Psi \ra,  \\
  e^{2\omega} \p_{-} \left[  e^{-2\omega}  \p_{-} \Phi \right] &=-8\pi G\; \la \Psi  | T_{--} |\Psi  \ra.
\end{split}
\label{eq:dilatoneom}
\end{equation}
We first fix the form of the stress tensor expectation value. Since the CFT is defined on a curved background,  
it acquires contributions from the Weyl anomaly,
\begin{equation} 
\la \Psi | T_{\pm \pm} |\Psi \ra  =\f{c}{12\pi} \left[\p^{2}_{\pm} \omega -(\p_{\pm} \omega)^{2} \right] + \tau_{\pm \pm} =\f{c}{48 \pi} +\tau_{\pm \pm},
\end{equation}
We have defined $\tau_{\pm \pm}$, which are the thermal expectation values of $T_{\pm\pm}$ on the strip with flat metric. 
For our choice of the state \eqref{eq:HHstate}, they are given by 
\be  
\tau_{\pm \pm} = \f{c}{24\pi} \left(\f{2\pi}{\beta} \right)^{2} - \f{c}{48 \pi} ,
\ee
The second term,  the Casimir energy arising from finite size effects, cancels the contribution of the Weyl factor in the total stress tensor expectation value. 
By plugging  these values, the solution of the equations of motion is given by 
\be 
\Phi (\tau, \mu )  =\Phi_{0}(\tau, \mu ) -\f{K}{2} \left( \mu \tan \mu +1 \right) - \f{cG}{3}, \label{eq:totaldilaton}
\ee
where $K$ is defined by
\be 
K=\f{4cG}{3} \left( \f{2\pi}{\beta}\right)^{2} .
\ee
In this solution, previously discussed in \cite{Bak:2018txn},
$\Phi_{0}(\tau, \mu )$ satisfies the equations of motions (\ref{eq:dilatoneom}) 
with $ \la T_{\mu \nu} \ra =0$, so we refer to it as a sourceless solution.  In the $(\tau, \mu) $ coordinates, we will choose
\be 
\Phi_{0}(\tau, \mu )  = \alpha_{0}\; \f{\cos \tau}{\cos \mu}. \label{eq;simpdil}
\ee

The sourceless dilaton profile \eqref{eq;simpdil} describes a two dimensional eternal black hole. In order to see this, let us first recall that the total dilaton $\phi_{0}+ \Phi$  represents the volume of the transverse directions of higher dimensional gravity, where $\phi_{0}$ is area of the horizon of the higher dimensional extremal black hole whose dimensional reduction gave the JT gravity (see \eqref{eq:ZB})  The zeros of the total dilaton thus correspond to curvature singularities of the spacetime. 
The total dilaton for $\Phi = \Phi_0$ vanishes at the AdS boundary $\mu=\f{\pi}{2}$ as soon as  $\tau$ gets larger than $\f{\pi}{2}$ or smaller than $-\f{\pi}{2}$.\footnote{Of course, if we have $\phi_0 > 0$, the Penrose diagram will be that of a maximally extended extremal AdS black hole (see \cite{Brown:2019rox} for a more detailed construction), but in this paper we will only be concerned with a single instance of the extension even for $\phi_0 > 0$.  So, we will continue to draw the Penrose diagram for $\phi_0>0$ with singularities at $\tau = \pm \frac{\pi}{2}$.}
Thus the spacetime singularities intersects with the boundary at $\tau= \pm \f{\pi}{2}$ 
The light rays starting from the intersections correspond to black hole horizons. 
In particular, the point $(\tau, \mu) =(0,0)$ corresponds to the bifurcation surface. 
This can also be seen from the fact that the derivatives of the dilaton vanish at this point, i.e. $\p_{\mu} \Phi_{0}(0,0) = \p_{\tau} \Phi_{0}(0,0) =0$, which is the 2d counterpart of the extremal surface condition in higher dimensions.

Next we want to fix the constant $\alpha_0$ in the sourceless dilaton profile \eqref{eq;simpdil} so that asymptotically $\mu \rightarrow \pm \f{\pi}{2}$,  the full solution \eqref{eq:totaldilaton} approaches an  AdS$_2$ black hole.  To do this, we recall from  \cite{Bak:2018txn} that the dilaton field
\begin{equation}
\Phi_{0}(\tau,\mu) = \f{\bar{\phi} L}{2} \left[ \left(b+\f{1}{b} \right) \f{\cos \tau}{\cos \mu} -\left(b-\f{1}{b} \right) \tan \mu\right], 
\label{eq:blackcaus} 
\end{equation}
together with the metric \eqref{eq:metric}, is a general solution of the sourceless equations of motion.
Moreover, by the coordinate transformations
\begin{equation}
\f{r}{L} = \f{ \left( b+ \f{1}{b}  \right) \cos \tau -\left( b- \f{1}{b}  \right) \sin \mu  }{2\cos \mu}, \quad
\tanh tL = \f{2\sin \tau }{\left( b+ \f{1}{b}  \right) \cos \tau -\left( b- \f{1}{b}  \right) \sin \mu},
\label{eq:coordtrans}
\end{equation}
this configuration can be mapped to
\begin{equation}
\Phi(t,r) =\bar{\phi} r, \quad
ds^{2} = -(r^{2} -L^{2} ) dt^{2} + \f{dr^{2}}{r^{2} -L^{2}}, \label{eq:ordinarybh}
\end{equation}
which describes a coordinate patch of and asymptotically AdS$_2$ black hole with temperature 
\be 
T=\f{L}{2\pi}. 
\label{eq:tempBH}
\ee
Notice that the new coordinates $(r,t)$ only cover the region outside the horizon. 

The dilaton profile \eqref{eq:blackcaus} depends on a parameter $b$. 
This parameter is related to the size of the black hole interior. To see this, let us first identify the location of the right event horizon by solving 
$\p_{\mu} \Phi_{0} = \p_{\tau} \Phi_{0} =0$ using \eqref{eq:blackcaus}. 
The $\tau$ equation implies $\tau = 0$, and the $\mu$ equation implies that the critical point $(\tau, \mu ) =(0, \mu_{0})$ satisfies
\be 
\sin \mu_{0} = \f{b^{2}-1}{b^{2}+1}.
\ee 
A similar condition can be obtained for the left black hole horizon, which will be located at $(\tau,\mu) = (0,-\mu_{0})$, by flipping the sign of $\tan \mu$ in the dilaton profile \eqref{eq:blackcaus}.
This creates a dilaton profile for the left asymptotic region, where $\mu$ is negative but we still require $r>L$.
When $b$ is large, the horizon approaches the AdS boundary, $\mu_{0} \rightarrow \f{\pi}{2}$.  The domain of dependence associated with the subregion $-\mu_{0} <\mu <\mu_{0}$ on the $\tau=0$ slice corresponds to what we refer to as the causal shadow region, which is never causally accessible from the asymptotic boundaries $\mu  = \pm \f{\pi}{2}$. 
It is easy to check that the dilaton at the left and right black hole horizons is independent of $b$ and has value $\bar{\phi} L$,  which is consistent with the fact that horizon exterior is a static black hole with the temperature \eqref{eq:tempBH}.
Therefore, the net effect of making $b\neq 1$ is to create a causal shadow region behind the horizon, while  the temperature and the entropy of the black hole remained fixed by $L$.

Having found the extremal surfaces, our last task with the sourceless solution \eqref{eq:blackcaus} is to enforce appropriate asymptotically AdS boundary conditions. This procedure will specify the free parameter $\alpha_{0}$ in \eqref{eq:totaldilaton}.  
We will demand that this dilaton describes a black hole with inverse temperature $\beta_{BH}$, for any entanglement temperature $\beta$.  
We can always do this by choosing $\alpha_{0}$ appropriately; we simply expand \eqref{eq:blackcaus} and \eqref{eq;simpdil} around $\mu \sim \frac{\pi}{2}$ and match the leading divergences.
This set up is quite similar to that of \cite{Penington:2019kki,Balasubramanian:2020hfs}, where radiation entropy was computed as a function of some entangling parameter ($k$ in \cite{Penington:2019kki}, $r_h'$ in \cite{Balasubramanian:2020hfs}) which increased, while keeping fixed the temperature of the black hole $\beta_{BH}$.  Here we have a fixed black hole in universe $B$, and we tune the amount of entanglement with the auxiliary universe $A$ by changing the parameter $\beta$.
 
\begin{figure}
\centering
\includegraphics[width=7cm]{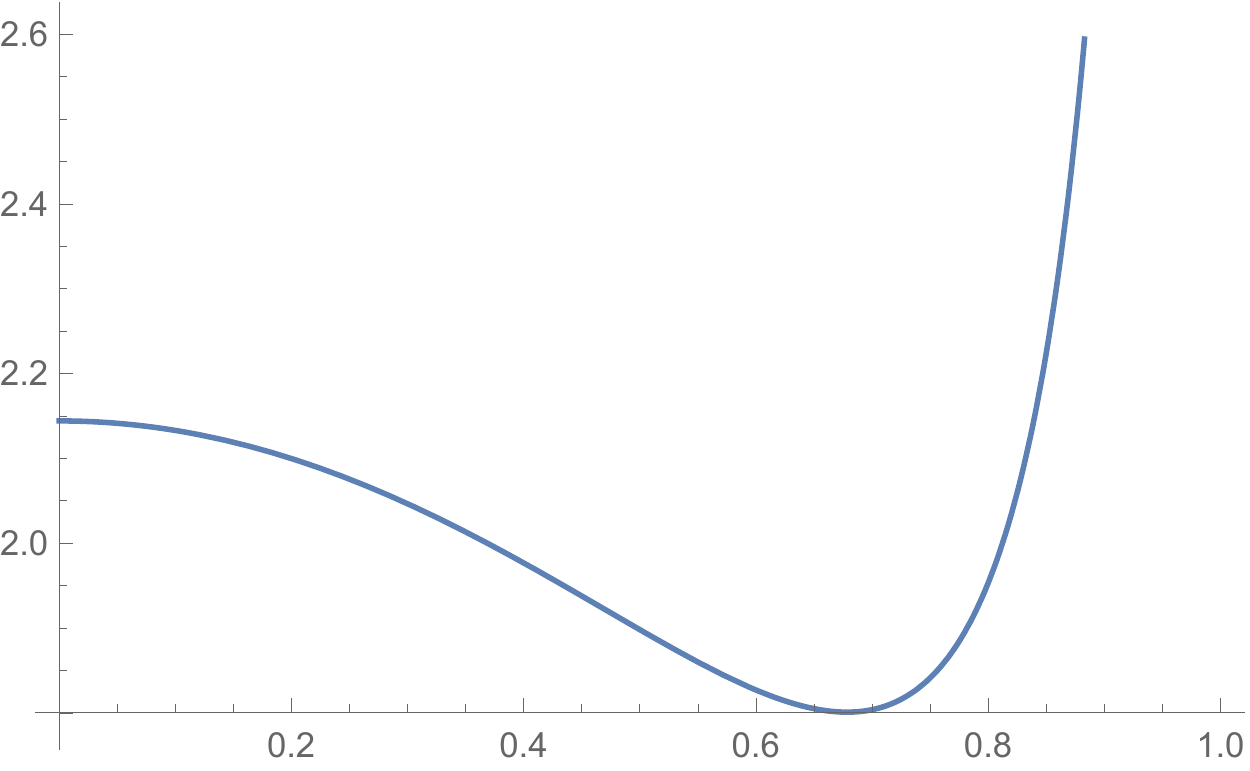}
\hspace{0.5cm}
\includegraphics[width=7cm]{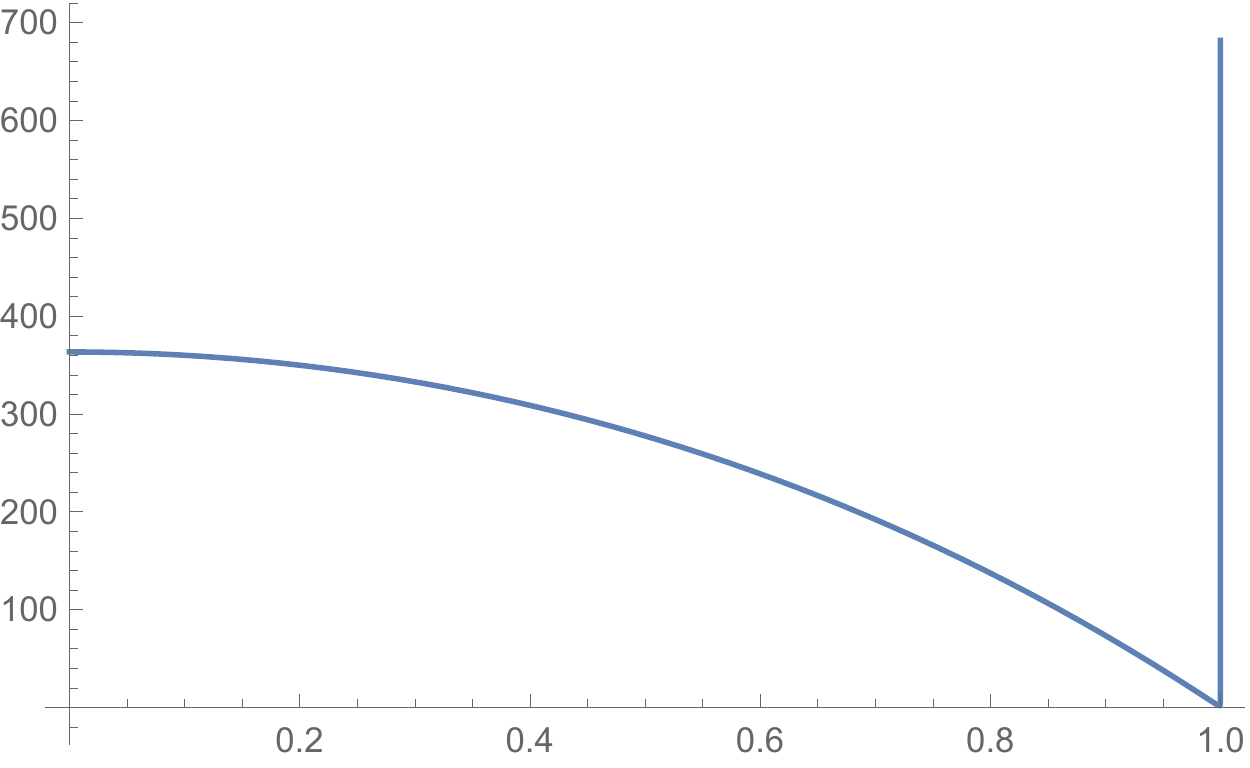}
\caption{\small{Plots of the dilaton profile \eqref{eq:totaldil} for $b=5$ (left) and $b=1000$ (right). Larger $b$ corresponds to higher temperature. The dilaton is minimized at the horizon, and its minimum value corresponds to the classical entropy of the black hole.}}
\label{figure:dilplot}
\end{figure}

We now apply this procedure to impose  asymptotic AdS boundary conditions which fix the black hole temperature as seen from infinity on the the solution with source \eqref{eq:totaldilaton}.  As described above, to do so we equate the leading divergences of \eqref{eq:blackcaus} and \eqref{eq:totaldilaton} (where in \eqref{eq:totaldilaton} we are using the global sourceless $\Phi_0$ defined in \eqref{eq;simpdil}) as $\mu \to \frac{\pi}{2}$.
This yields the following relationship between $b,\bar{\phi}$ and $\alpha_0,K$:
\be 
\f{\pi K}{4} = \f{\bar{\phi}L_{\beta_{BH}}}{2} \left(b_{0}-\f{1}{b_{0}} \right)  , \quad \alpha_{0} =\f{\bar{\phi}L_{\beta_{BH}}}{2} \left(b_{0}+\f{1}{b_{0}} \right) , 
\label{eq:betacod}
\ee 
where we have defined
\begin{equation}
L_{\beta_{BH}} =\f{2\pi}{\beta_{BH}} .
\end{equation}
\begin{figure}[t]
    \centering
    \includegraphics[scale=.3]{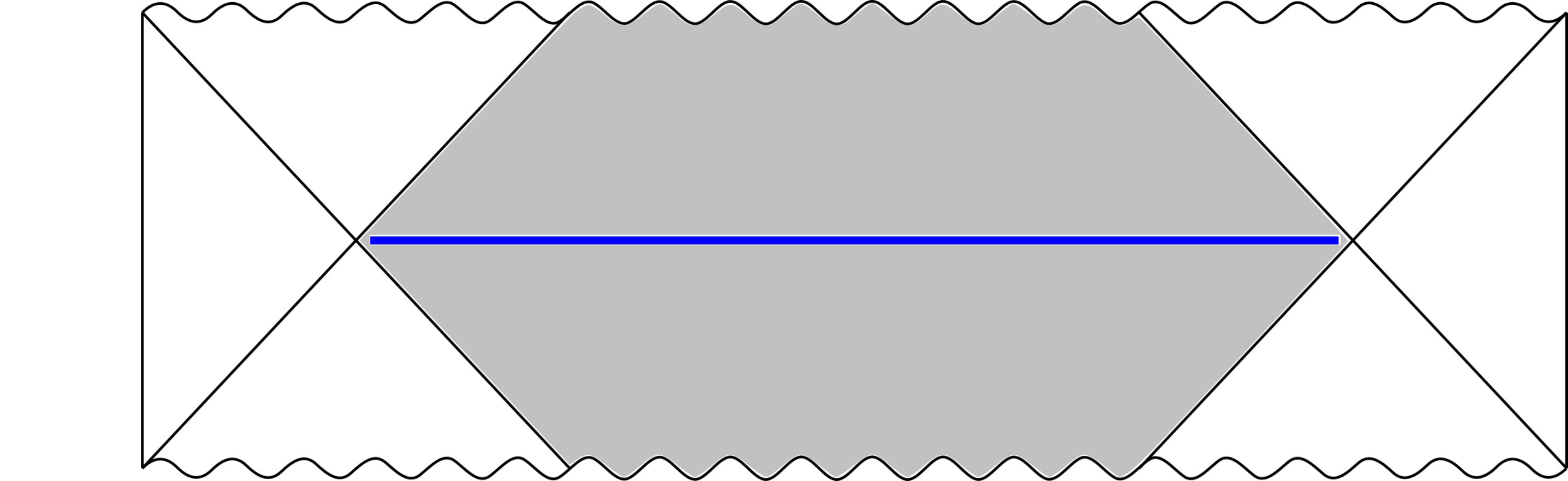}
    \caption{\small{The Penrose diagram of the backreacted black hole with the dilaton profile \eqref{eq:totaldil}. Near asymptotic boundaries $\mu \rightarrow \f{\pi}{2}$,  it looks like an eternal black hole \eqref{eq:ordinarybh} with a fixed temperature $1/\beta_{BH}$.   The shaded region is the causal shadow region, which is not causally accessible from both of the asymptotic  boundaries.  As we increase the entanglement temperature $1/\beta $, the size of the causal shadow region 
increases, and the event horizons approach the boundaries. The blue line in the $\tau=0$ slice is the island of the black hole.
 }}
    \label{fig:causalsh}
\end{figure}
Thus, the final form of our dilaton is
\be 
\Phi(\tau,\mu) = \f{\bar{\phi}L_{\beta_{BH}}}{2} \left(b_{0}+\f{1}{b_{0}} \right) \f{\cos \tau}{\cos \mu }  -\f{\bar{\phi}L_{\beta_{BH}}}{\pi} \left(b_{0}-\f{1}{b_{0}}  \right) \left( \mu \tan \mu +  1 \right) -\f{cG}{3}. \label{eq:totaldil}
\ee
Although the black hole temperature is fixed, the location of the 
of the black hole horizon, determined now by extrema of $\Phi$,  depends on the entanglement temperature $1/\beta$  Since $b_{0}$ satisfies \eqref{eq:betacod},  $b_{0} \sim 1/\beta^{2} $ in the $\beta \rightarrow 0$ limit. This implies that as we increase the entanglement temperature $1/\beta$, the causal shadow region is growing, and therefore the horizons of the black hole are approaching the AdS boundaries. 
To visualize this, we plot the dilaton profile as a function of $b$ in Fig.~\ref{figure:dilplot}.

\subsection{Minimization of generalized entropy} 

Having specified the  dilaton profile \eqref{eq:totaldil}, we can perform the minimization of the generalized entropy. 
We will take the following symmetric ansatz for the island in  $(\tau, \mu)$ coordinates:
\be 
 C_x:\quad \tau=0, \quad  - \f{\pi }{2} x <\mu <\f{\pi}{2} x,  \quad  0<x<1. \label{eq:theregion}
\ee
Since the dilaton profile is symmetric under $\mu \to -\mu$, the generalized entropy for this ansatz is given by 
\be 
S_{\text{gen}} (x) =2\phi( x) + S_\beta[\overline{C_x}] - S_{{\rm vac}}[\overline{C_x}]  ,  \quad \phi(x) \equiv \phi\left(\tau=0,\; \mu = \f{\pi}{2}x \right) , \label{eq:genen2}
\ee
where the dilaton profile is given by \eqref{eq:totaldil}. The second term is the CFT entanglement entropy of the complement of the island region \eqref{eq:theregion}, which at large $c$ takes the form
\be 
S_\beta[\overline{C_x}] - S_{{\rm vac}}[\overline{C_x}] = \f{c}{3} \log \left[ \f{\beta}{\pi} \sinh \left(\f{\pi^{2}(1-x) }{\beta} \right) \right] - \f{c}{3}\log 2\sin \pi x .
\ee 
Before doing this minimization in detail, we first argue that the emergence of a nontrivial island is robust, i.e. it does not depends on the details of the actual dilaton profile. 
This can be seen from the asymptotic behavior of the generalized entropy.  
When the size of the interval is small, $x \rightarrow 0$, the CFT entanglement entropy term is large,
\be
S_\beta[\overline{C_x}] - S_{{\rm vac}}[\overline{C_x}]   \rightarrow 
 \f{c}{3} \log \left[\f{\beta}{\pi} \sinh \left(\f{\pi^{2} }{\beta} \right)\right] -\f{c}{3}\log  x  \gg 1, \quad x \rightarrow 0. 
 \ee
By contrast, in the opposite limit, the dilaton $\Phi(x)$ diverges due to the boundary condition
\be
\Phi(x)  \rightarrow \f{A}{1-x}, \quad x \rightarrow 1 ,
\ee
with some coefficient $A$. 
Thus, the generalized entropy has to have a minimum in the middle, between $x \to 0$ and $x \to 1$.
Of course, this argument only implies the emergence of an extremal island, but does not imply that it has a smaller entropy than the CFT thermal entropy of universe $A$, which is necessary for the island to dominate the entropy calculation.

\begin{figure}
    \centering
    \includegraphics[height=0.4\textwidth]{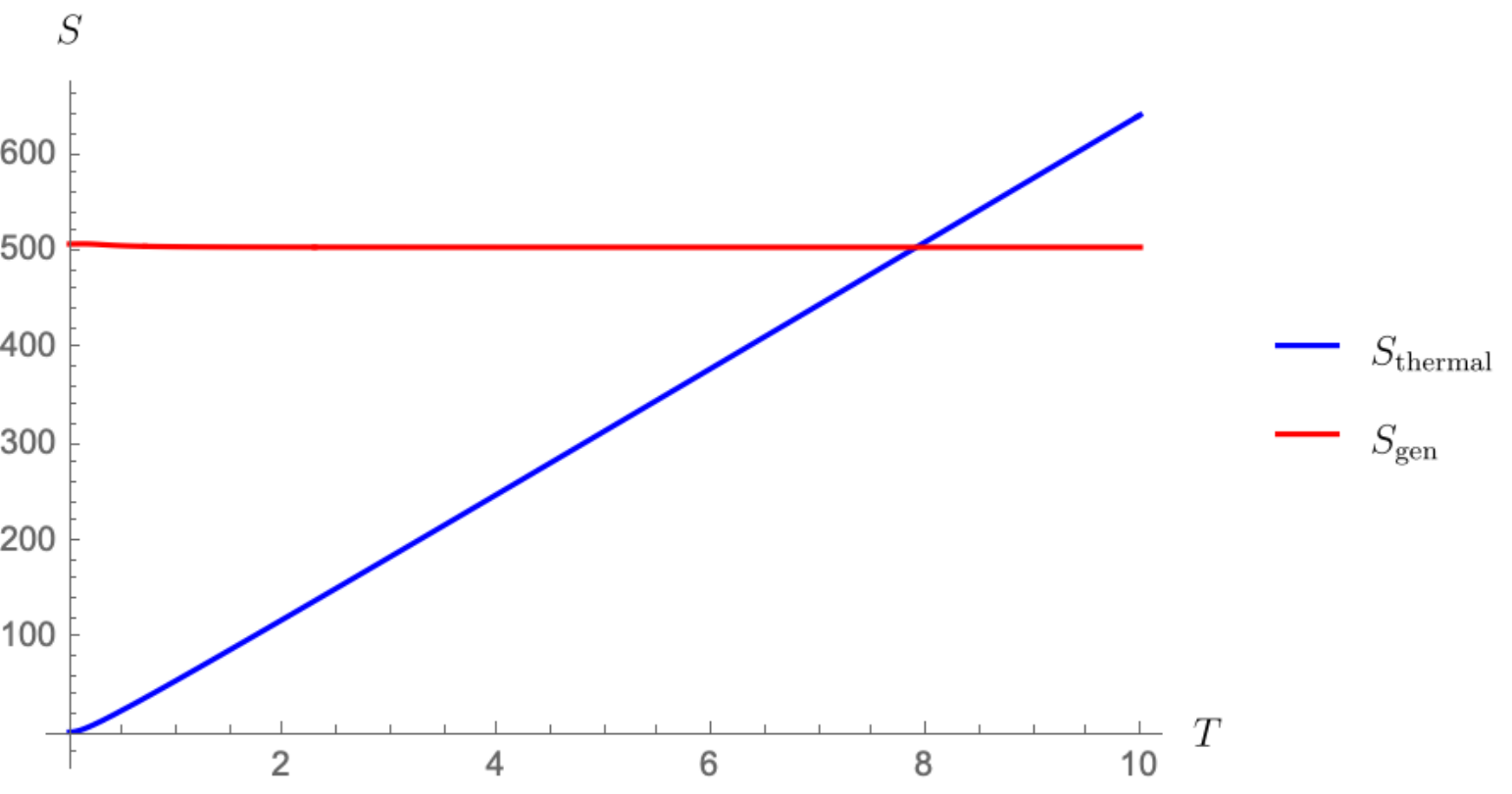}
    \caption{\small{A plot of the two candidate entropies for $S_A$ as a function of the CFT temperature $T = 1/\beta$, with $\phi_0 = 1000$, $\phi_b = c = 10$, and $L = G_N  =1$.  The blue curve is the thermal entropy of CFT fields on universe $A$, and corresponds to dominance of the disconnected saddle in the replica trick computation of the entropy.  The red curve is the generalized entropy of a complement island on universe $B$, which corresponds to dominance of the fully connected replica wormhole.  The Page curve in AdS (the entropy $S_A$) is computed by taking the minimum of these two curves.  We see that the Page transition occurs around $T = 8$.}}
    \label{fig:ads-page}
\end{figure}

With this remark in mind, let us now minimize the generalized entropy \eqref{eq:genen2} and find the quantum extremal surface. 
We have shown that the backreaction caused by the CFT entanglement creates a large causal shadow region in the black hole interior.
This pushes the classical horizons to the boundary.
Therefore, the quantum extremal surface also approaches the AdS boundary, for large enough $1/\beta$. 
This is because in this limit the contribution of the field theory entropy, as well as its derivative, vanishes:
\be 
S_\beta[\overline{C_x}] - S_{{\rm vac}}[\overline{C_x}] \sim  \f{c}{3} \left[ \log (1 - x) -\log \sin \pi x \right], \quad x  \rightarrow 1.
\ee
Thus, the quantum extremal surface almost coincides with the classical horizon of the black hole. 
Therefore, in this limit the contribution to the entropy of the connected wormhole is given by the area of the classical horizon, $S_{{\rm gen }} =2\pi \bar{\phi} /\beta_{BH}$.
Finally, we need to minimize over all saddles. 
The fully disconnected saddle, which gives the Hawking result \eqref{eq:QFTresult}, yields $\; S_{{\rm disconnected}} =2\pi^{2}c/3\beta$ in the high temperature limit.
In this limit, the fully connected saddle (as we have just discussed) gives the generalized entropy $S_{{\rm gen }} =2\pi \bar{\phi} /\beta_{BH}$.
The actual entropy is the minimum of these. 
Since $\beta_{BH}$ is constant and we are increasing $1/\beta$, we see that the Page transition occurs when the temperature of the radiation (or, equivalently, the entanglement temperature) becomes sufficiently larger than that of the black hole. 
Our analytic discussion has been in the high temperature limit, but we can  numerically minimize the entropy throughout the temperature range (Fig.~\ref{fig:ads-page}).

Note that a nontrivial island forms only when the original state is entangled. For example, suppose we start from the unentangled state $ | W \ra_{A} \otimes  | V \ra_{B}$, instead of \eqref{eq:HHstate}.  
A nontrivial island cannot form in this case.
In order to see this, suppose there is a replica wormhole solution which induces a cut of size $2\pi x$ after the replica symmetry quotient.
Then a similar calculation to the one discussed in Sec.~\ref{sec:setup} yields  
\be
S(\rho_{A}) = \min_{x} \left[  \phi(x) + S_{V} [\overline{C_x}] - S_{{\rm vac}}  [\overline{C_x}]\right] ,
\ee
where $S_{V} [\overline{C_x}] $ is the entanglement entropy of the pure state $| V \ra_{B}$ on the interval $\overline{C_x}$ of the size $2\pi (1-x)$. 
The dilaton profile will depend on the particular states we have chosen through the stress energy tensor; this is not so different from what we discussed earlier.
However, the CFT entropy behaves very differently, since the reduced density matrix on the universe $B$ is no longer mixed. In particular, it satisfies the pure state relation $S_{V}[\overline{C_x}] =S_{V}[C_{x}]$. 
As a consequence, the CFT entropy contribution no longer diverges as the wormhole becomes small $x \to 0$; rather, it goes to zero in this limit.
Therefore, extremizing over complement islands tells us that the complement of the extremal island is the whole Cauchy slice of $B$, and therefore there is no island.
Note that, though the island formula has given us the correct answer, a replica wormhole treatment of this pure state situation will need to take into account important subtleties in the replica trick \cite{Engelhardt:2020qpv} to ensure we get identically zero for the pure state entropy.

\section{Discussion}
\label{eq:discussion}

\subsection{Classical correlations do not produce islands}

We can 
demonstrate that density matrices on the bipartite Hilbert space which only contain classical correlations cannot generate islands. 
We start from the so-called thermo-mixed double (TMD) state \cite{Verlinde:2020upt}
\be 
\rho^{\text{TMD}}= \sum_{i}p_{i}\; |i \ra \la i|_A \otimes |\psi_{i} \ra \la \psi_{i}|_B ,
\ee
which is separable, and therefore contains no quantum entanglement. 
We would like to know whether or not a nontrivial island emerges in the calculation of $S_A$ in this state. 
The coefficients $p_{i}$ are again the Boltzmann factors defined in \eqref{eq:HHstate}.
The R\'enyi entropy of the TMD is given by 
\be
{\rm tr} \; (\rho^{\text{TMD}}_{A})^{n} =\sum_{i}p_{i}^{n} \; ( \la \psi_{i} | \psi_{i}  \ra_{B})^{n}.
\label{eq:TMD-renyi}
\ee
This can be compared with the result \eqref{eq:renyin} for the pure state \eqref{eq:HHstate}. The R\'enyi entropy for the pure state involves overlaps of different states $\la \psi_{i}| \psi_{j}\ra_{B}$, whereas the R\'enyi entropy in \eqref{eq:TMD-renyi} involves overlaps of the identical states. 
In quantum field theory, this R\'enyi entropy is given by a ratio of partition sums
\be 
{\rm tr} \; (\rho^{\text{TMD}}_{A})^{n}  \supset \f{Z(n\beta)}{Z^{n}(\beta)} \, ,
\ee
which is identical to the pure state result \eqref{eq:QFTresult}.
The difference between the pure state \eqref{eq:HHstate} and the TMD appears only when we take into account the gravitational effects in universe $B$. 
Here again we will assume that the leading gravitational effect comes from a fully connected wormhole which joins all copies of universe $B$ with the metric \eqref{eq:metric}.  
The product of overlaps in this case is given by
\be 
(\la \psi_{i} | \psi_{i}  \ra_{B})^{n} =\f{ {\rm tr} (\rho_{\psi_{i}})^{n}}{{\rm tr} (\rho_{\psi_{{\rm vac}}})^{n}}.
\ee
So, the CFT component of the connected saddle gives a vacuum-normalized excited state R\'enyi entropy, which of course depends on the particular CFT operator $\psi_i$.
Furthermore, we again take the ansatz \eqref{eq:theregion} for the island on the AdS$_2$ strip.
The contribution of the wormhole to the entanglement entropy in this case is given by 
\be 
S_{\text{conn}}= S_\beta (B) + \underset{x}{\text{min ext}} \left[ 2\phi[\partial \overline{C_x}] + \sum_{i} p_{i} \left(S_{\rho_{\psi_i}}[\overline{C_x}] -S_{{\rm vac}}[\overline{C_x}] \right) \right] , \label{eq:location}
\ee
where $S_\beta (B) =-\sum_{i} p_{i} \log p_{i}$ is the thermal entropy. 
The dilaton profile is still given by \eqref{eq:totaldil}, as the total expectation value of the stress energy tensor remains unchanged, since the sum of the individual dilatons associated to the pure states (together with the weightings $p_i$) recombine to form the thermal stress energy; this is simply because the dilaton equation of motion is linear.

Finally, we need to minimize over $x$ and compare with $S_\beta (B)$.
If there is a Page transition and an island phase, we will see it in the high temperature limit $\beta \rightarrow 0$.
In this limit, \eqref{eq:location} is minimized near the AdS boundary, when $x \to 1$. This is because the dilaton contribution is much larger than the CFT entropy contribution, and thus the function is minimized at the classical horizon of the black hole which is located near the boundary.
As a consistency check when $x \rightarrow 1$, $S_{\rho_{i}}[\bar{C}_{x}] -S_{{\rm vac}}[\bar{C}_{x}]$ vanishes for any $\rho_{\psi_i}$.
Therefore, we conclude $S_{\text{conn}}= S_\beta (B) +S_{BH} \gg S_\beta (B)$. This implies that no island appears in the calculation of the entropy of the TMD state, which only contains classical correlations.  This confirms that
the appearance of the island is a consequence of the quantum nature of the information which is stored in the entanglement between the two universes, consistent with the idea that quantum entanglement leads to spacetime connectivity \cite{VanRaamsdonk:2010pw}.  

At first glance, our finding  seems to be in tension with the result in \cite{Verlinde:2020upt} that the correlations associated with connectedness of space across an Einstein-Rosen bridge can be mostly classical with the quantum entanglement restricted to a code subspace.  Similarly see \cite{Balasubramanian:2014gla} for results suggesting that strong entanglement does not have to create semiclassical wormholes.   However, like in \cite{Verlinde:2020upt} we are effectively considering entanglement within a low-energy code subspace and that seems sufficient to require the inclusion of the Euclidean wormholes in the path integral that produce the island phenomenon.  It would be nice to better understand the relationship between these results.

\subsection{A holographic description}

\begin{figure}[t]
    \centering
    \includegraphics[scale=.3]{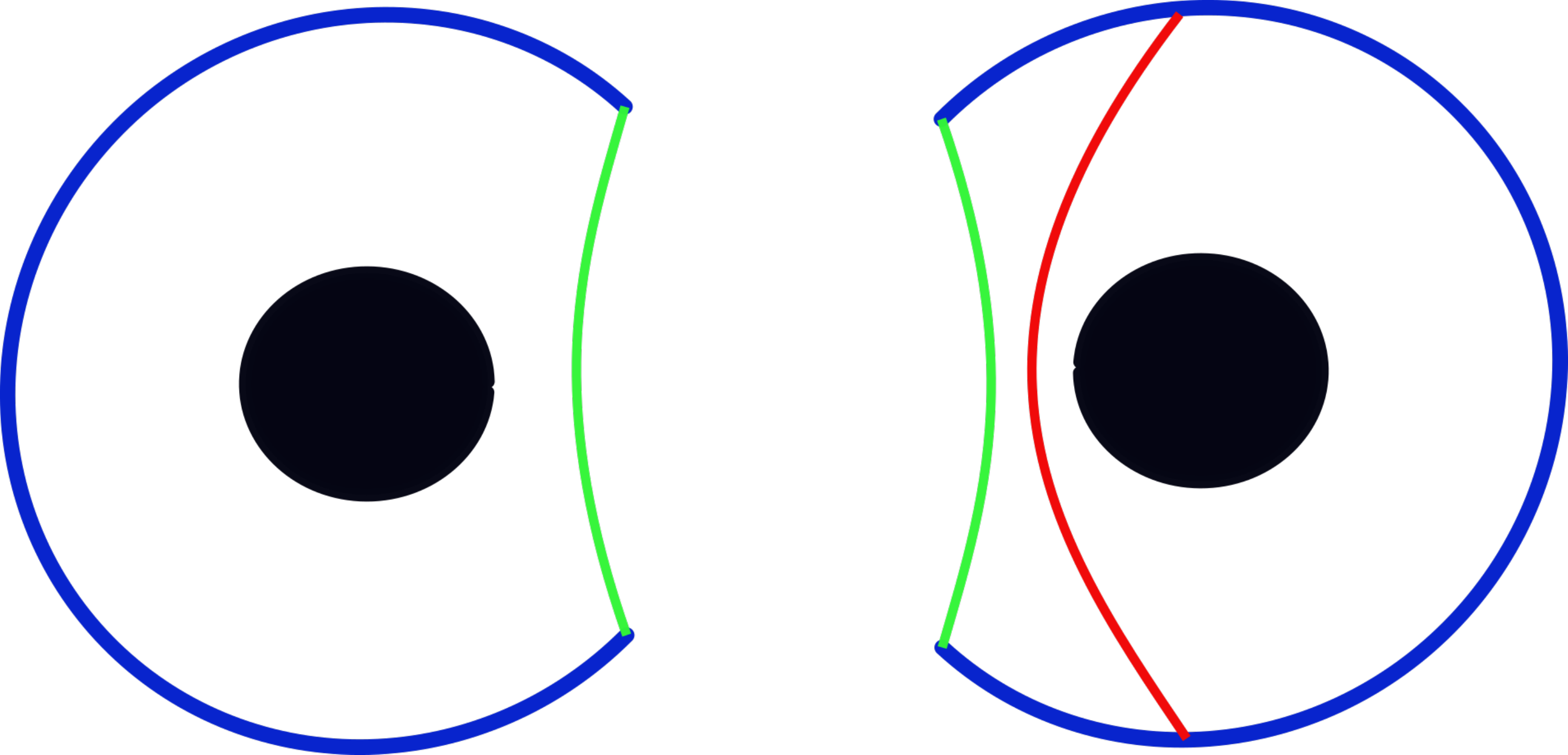}
    \caption{\small{A time slice of three dimensional  holographic dual  of the thermofield double state on two disjoint universes (a BTZ black hole). We choose each universe to be $AdS_{2}$ (blue intervals). CFT on this segment is dual to the bulk region surrounded by the segment and the probe brane (green line) \cite{Takayanagi:2011zk}.  In the right gravitating universe we have a black hole. The red line is the RT surface which ends on the two horizons of the boundary black hole.}}
    \label{fig:holographic}
\end{figure}

If we choose the CFT on universes $A$ and $B$ to be holographic, we can construct a  dual description of our system using classical gravity in AdS$_3$. 
Then, the  state \eqref{eq:HHstate} is a thermofield double, which is dual to a two-sided eternal BTZ black hole with boundary geometry that matches the geometry of universes $A$ and $B$ (see Fig.~\ref{fig:holographic}).
Since both universes have a metric which is not flat, the boundary surfaces on which they live are not quite asymptotic, but are instead slightly pushed into the bulk.
The precise location of these boundary surfaces is determined by the Weyl factor of the corresponding 2d boundary metric.

In this setting, we can use holography to calculate the CFT part of the generalized entropy by using the Ryu-Takayanagi (RT) formula.  We saw above that this term reduces to the CFT entanglement entropy of the complement of the Euclidean wormhole region in universe $B$.  This is because the cut producing the wormhole is identified with the island in our setup. In the holographic setting,
the same CFT entanglement entropy  can be computed by the length of the geodesic in the BTZ black hole connecting the two end points of the complement region in universe $B$. 

Let us consider the effect of increasing the entanglement temperature $1/\beta$ in \eqref{eq:HHstate}.
This will: (1) increase the size of the bulk BTZ black hole, and (2) increase the size of the island in the black hole interior in the boundary surface corresponding to universe $B$. As the island size increases, its complement shrinks, and the RT surface approaches the asymptotic boundary.  This implies that the entanglement wedge of universe $A$ grows, and eventually covers the whole bulk spacetime in the high temperature limit.  In particular, we can eventually reconstruct the whole 3d bulk just from the quantum information in universe $A$.  This statement can be regarded as a refinement of the results of \cite{Balasubramanian:2020hfs}, where the interior horizon structure of a 3d bulk prevented total reconstruction in the radiation regions.

\subsection*{Acknowledgments}

We thank Norihiro Iizuka, Yuki Miyashita, Simon F.~Ross, Masaki Shigemori, Tadashi Takayanagi, and Kotaro Tamaoka for useful discussions.
VB and AK were supported in part by the Simons Foundation through the It From Qubit Collaboration (Grant No.~38559), and by the Department of Energy through grants DE-SC0013528, and QuantISED  DE-SC0020360.
 TU was supported by JSPS Grant-in-Aid for Young Scientists  19K14716. VB also thanks the Aspen Center for Physics, which is supported by National Science Foundation grant PHY-1607611.

\appendix

\section{Proof of replica path integral identity}\label{sec:identity}

\begin{figure}[t]
    \centering
    \includegraphics[scale=.5]{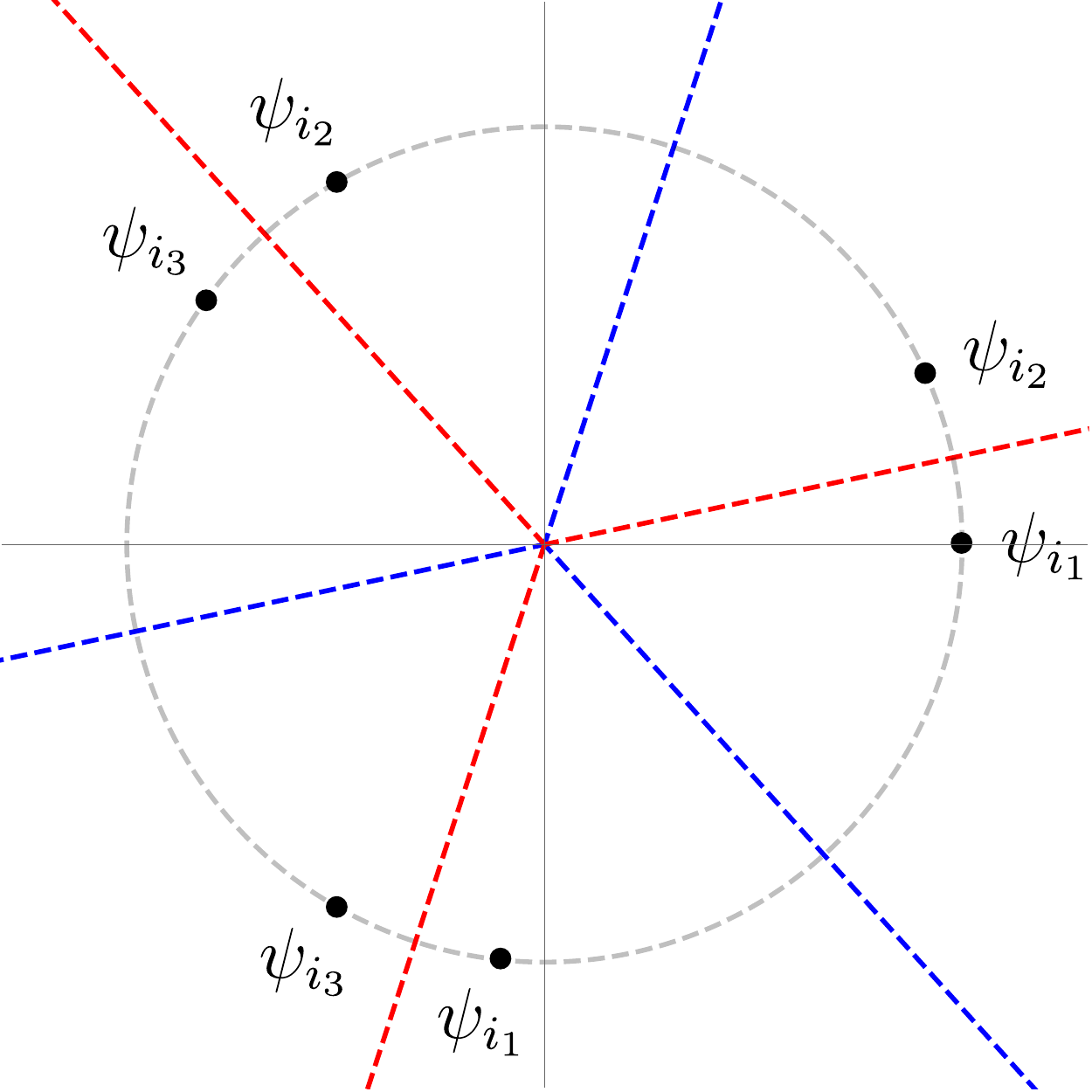}
    \caption{\small{The uniformized plane for $n=3$, $x=0.2$.  Branch cuts from the $\Sigma_{3}(C_x)$ manifold are shown as dashed blue lines, and branch cuts from $\Sigma_{3}(\overline{C_x})$ are shown as dashed red lines.  These cuts are the uniformized images of the cuts appearing in Fig.~\ref{fig:sphere-replicas}.  Notice that the blue cut images separate operators of the same type, and the red cut images separate operators of different type.  This allows for an equivalence between the branched sphere correlators for the two patterns of operator insertions and cuts.}}
    \label{fig:uniformized-sphere}
\end{figure}

In the body of the paper, we used the identity  \eqref{eq:corridentity} which we reproduce here:
\begin{equation}
    \avg{\psi_{i_1}(\infty_1) \psi_{i_2}(0_1) \dots \psi_{i_n}(\infty_n) \psi_{i_1}(0_n)}_{\Sigma_{n}(C_x)} = \avg{\psi_{i_1}(\infty_1) \psi_{i_1}(0_1) \dots \psi_{i_n}(\infty_n) \psi_{i_n}(0_n)}_{\Sigma_{n}(\overline{C_x})} .
    \label{eq:correpationapp}
\end{equation}
$\Sigma_{n} (C_{x})$ in the above expression is constructed by starting from the two dimensional sphere  
\be 
 ds^{2} =d\theta^{2} +\sin^{2} \theta d \phi^{2},  
\ee
and gluing $n$ copies of the sphere cyclically along the cut $C_{x}:\theta=\f{\pi}{2}, \;  0<\phi< 2\pi x$.  
$\Sigma_{n} (\overline{C_x})$ is again an $n$-branched sphere with the cut  $\overline{C_x}: \theta=\f{\pi}{2},\;  2\pi x <\phi< 2\pi$.  
It is convenient to first map the sphere to the cylinder $(t, \theta)$ by 
\be
t =\log \tan \f{\theta}{2}.
\ee
The location of the cut is $t=0,0<\phi< 2\pi x $.
This branched cylinder is mapped to a branched plane by $z=e^{i \theta +t}$.
We then uniformize  the branched plane  by the conformal map \cite{Sarosi:2016oks,Asplund:2014coa}
\be 
w=\left( \f{z-e^{2\pi i x}}{z-1}\right)^{\f{1}{n}} .
\ee
Then, each operator location is mapped to
\be 
z=\infty_{k} \rightarrow w_{k}=  e^{\f{2\pi i k}{n}}, \quad z=0_{k} \rightarrow \hat{w}_{k} =e^{\f{2\pi i (k+x)}{n}} .
\ee
This allows us to write the left hand side of \eqref{eq:correpationapp}  in terms of a correlation function $\la \cdot \ra_{\mathbb{C}}$ on the plane $\mathbb{C}$,
\be
\la \prod_{k=1}^{n}  \psi_{i_{k}} (\infty_{k})  \psi_{i_{k+1}} (0_{k})  \ra_{\Sigma_{n}(C_x)} = J(x) \; \la \prod_{k=1}^{n}  \psi_{i_{k}} (w_{k})  \psi_{i_{k+1}} (\hat{w}_{k} ) \ra_{\mathbb{C}} ,
\ee
with a Jacobian factor $J(x)$ whose details are  found in \cite{Sarosi:2016oks}.
On the other hand, if we use the uniformization map
\be 
w=\left( \f{z-e^{-2\pi i(1-x)}}{z-1}\right)^{\f{1}{n}} ,
\ee
which maps the branched sphere insertions to
\be 
z= \infty_{k} \rightarrow w_{k}=  e^{\f{2\pi i k}{n}}, \quad z=0_{k} \rightarrow \hat{w}_{k-1}  = e^{\f{2\pi i (k-1+x) }{n}} ,
\ee
we find that the same correlation function may be expressed as
\be
\la \prod_{k=1}^{n} \psi_{i_{k}} (\infty_{k}) \psi_{i_{k}} (0_{k}) \ra_{\Sigma_{n}(\overline{C_x})} = J(x)\; \la \prod_{k=1}^{n}  \psi_{i_{k}} (w_{k})  \psi_{i_{k+1}} (\hat{w}_{k} ) \ra_{\mathbb{C}} .
\ee
Therefore, the claimed identity holds. 
The two methods of uniformization lead to Fig.~\ref{fig:uniformized-sphere} in the $n=3$ case.

\section{Dominance of the fully connected wormhole in high temperature limit}

In this appendix, we show that for sufficiently high temperature, the fully connected wormhole will dominate over other less connected wormholes.
In order to see this, 
let us consider the profile  where first $m$ universes are connected by a wormhole, and the rest are connected by another one.   The contribution of such configuration is given by 
\be 
I_{m, n-m} =\sum_{i_{1} \cdots i_{n}} p_{i_{1} } \cdots p_{i_{n} }  \la \psi_{i_{1}} \cdots \psi_{i_{m+1}} \ra_{\Sigma_{m}} \la \psi_{i_{m+1}}  \cdots \psi_{i_{1}}  \ra_{\Sigma_{n-m}}. 
\ee
The contribution gets maximized when  the size of both wormholes is maximal, $x=1$. However, even in this case, $I_{m, n-m} =\f{Z_{2\beta}}{Z_{\beta}^{2}}$ which is highly suppressed in the 
high temperature limit.  One can also check that other wormhole profiles give smaller contributions.  

One the other hand, the contribution of fully connected  wormhole is given by 
\be 
\sum_{i_{1} \cdots i_{n}} p_{i_{1} } \cdots p_{i_{n} }   \la \psi_{i_{1}} \cdots \psi_{i_{n}} \psi_{i_{1}}  \ra_{\Sigma_{n}} \rightarrow \sum_{i_{1} \cdots i_{n}} p_{i_{1} } \cdots p_{i_{n} }   \prod_{m=1}^{n} \la \psi_{i_{m}} (\infty_{m} )\psi_{i_{m}} (0_{m})\ra_{\Sigma_{1}} =1, 
\ee
in the large wormhole limit $x \rightarrow 1$, because this corresponds to the OPE channel $\psi_{i_{m}} (\infty_{m} ) \rightarrow \psi_{i_{m}} (0_{m} ) $.  Therefore this is  the unique contribution which remains finite in  the high temperature limit.

\bibliographystyle{JHEP}
\bibliography{renyi}

\providecommand{\href}[2]{#2}\begingroup\raggedright\begin{thebibliography}{10}

\bibitem{Almheiri:2019hni}
A.~Almheiri, R.~Mahajan, J.~Maldacena and Y.~Zhao, \emph{{The Page curve of
  Hawking radiation from semiclassical geometry}},
  \href{http://dx.doi.org/10.1007/JHEP03(2020)149}{\emph{JHEP} {\bf 03} (2020)
  149}, [\href{https://arxiv.org/abs/1908.10996}{{\tt 1908.10996}}].

\bibitem{Ryu:2006bv}
S.~Ryu and T.~Takayanagi, \emph{{Holographic derivation of entanglement entropy
  from AdS/CFT}},
  \href{http://dx.doi.org/10.1103/PhysRevLett.96.181602}{\emph{Phys. Rev.
  Lett.} {\bf 96} (2006) 181602},
  [\href{https://arxiv.org/abs/hep-th/0603001}{{\tt hep-th/0603001}}].

\bibitem{Ryu:2006ef}
S.~Ryu and T.~Takayanagi, \emph{{Aspects of Holographic Entanglement Entropy}},
  \href{http://dx.doi.org/10.1088/1126-6708/2006/08/045}{\emph{JHEP} {\bf 08}
  (2006) 045}, [\href{https://arxiv.org/abs/hep-th/0605073}{{\tt
  hep-th/0605073}}].

\bibitem{Hubeny:2007xt}
V.~E. Hubeny, M.~Rangamani and T.~Takayanagi, \emph{{A Covariant holographic
  entanglement entropy proposal}},
  \href{http://dx.doi.org/10.1088/1126-6708/2007/07/062}{\emph{JHEP} {\bf 07}
  (2007) 062}, [\href{https://arxiv.org/abs/0705.0016}{{\tt 0705.0016}}].

\bibitem{Faulkner:2013ana}
T.~Faulkner, A.~Lewkowycz and J.~Maldacena, \emph{{Quantum corrections to
  holographic entanglement entropy}},
  \href{http://dx.doi.org/10.1007/JHEP11(2013)074}{\emph{JHEP} {\bf 11} (2013)
  074}, [\href{https://arxiv.org/abs/1307.2892}{{\tt 1307.2892}}].

\bibitem{Engelhardt:2014gca}
N.~Engelhardt and A.~C. Wall, \emph{{Quantum Extremal Surfaces: Holographic
  Entanglement Entropy beyond the Classical Regime}},
  \href{http://dx.doi.org/10.1007/JHEP01(2015)073}{\emph{JHEP} {\bf 01} (2015)
  073}, [\href{https://arxiv.org/abs/1408.3203}{{\tt 1408.3203}}].

\bibitem{Almheiri:2019psf}
A.~Almheiri, N.~Engelhardt, D.~Marolf and H.~Maxfield, \emph{{The entropy of
  bulk quantum fields and the entanglement wedge of an evaporating black
  hole}}, \href{http://dx.doi.org/10.1007/JHEP12(2019)063}{\emph{JHEP} {\bf 12}
  (2019) 063}, [\href{https://arxiv.org/abs/1905.08762}{{\tt 1905.08762}}].

\bibitem{Penington:2019npb}
G.~Penington, \emph{{Entanglement Wedge Reconstruction and the Information
  Paradox}},  \href{https://arxiv.org/abs/1905.08255}{{\tt 1905.08255}}.

\bibitem{Penington:2019kki}
G.~Penington, S.~H. Shenker, D.~Stanford and Z.~Yang, \emph{{Replica wormholes
  and the black hole interior}},  \href{https://arxiv.org/abs/1911.11977}{{\tt
  1911.11977}}.

\bibitem{Almheiri:2019qdq}
A.~Almheiri, T.~Hartman, J.~Maldacena, E.~Shaghoulian and A.~Tajdini,
  \emph{{Replica Wormholes and the Entropy of Hawking Radiation}},
  \href{https://arxiv.org/abs/1911.12333}{{\tt 1911.12333}}.

\bibitem{Page:1993wv}
D.~N. Page, \emph{{Information in black hole radiation}},
  \href{http://dx.doi.org/10.1103/PhysRevLett.71.3743}{\emph{Phys. Rev. Lett.}
  {\bf 71} (1993) 3743--3746},
  [\href{https://arxiv.org/abs/hep-th/9306083}{{\tt hep-th/9306083}}].

\bibitem{Anegawa:2020ezn}
T.~Anegawa and N.~Iizuka, \emph{{Notes on islands in asymptotically flat 2d
  dilaton black holes}},  \href{https://arxiv.org/abs/2004.01601}{{\tt
  2004.01601}}.

\bibitem{Hashimoto:2020cas}
K.~Hashimoto, N.~Iizuka and Y.~Matsuo, \emph{{Islands in Schwarzschild black
  holes}}, \href{http://dx.doi.org/10.1007/JHEP06(2020)085}{\emph{JHEP} {\bf
  06} (2020) 085}, [\href{https://arxiv.org/abs/2004.05863}{{\tt 2004.05863}}].

\bibitem{Gautason:2020tmk}
F.~F. Gautason, L.~Schneiderbauer, W.~Sybesma and L.~Thorlacius, \emph{{Page
  Curve for an Evaporating Black Hole}},
  \href{http://dx.doi.org/10.1007/JHEP05(2020)091}{\emph{JHEP} {\bf 05} (2020)
  091}, [\href{https://arxiv.org/abs/2004.00598}{{\tt 2004.00598}}].

\bibitem{Krishnan:2020oun}
C.~Krishnan, V.~Patil and J.~Pereira, \emph{{Page Curve and the Information
  Paradox in Flat Space}},  \href{https://arxiv.org/abs/2005.02993}{{\tt
  2005.02993}}.

\bibitem{Hartman:2020swn}
T.~Hartman, E.~Shaghoulian and A.~Strominger, \emph{{Islands in Asymptotically
  Flat 2D Gravity}},  \href{https://arxiv.org/abs/2004.13857}{{\tt
  2004.13857}}.

\bibitem{Dong:2020uxp}
X.~Dong, X.-L. Qi, Z.~Shangnan and Z.~Yang, \emph{{Effective entropy of quantum
  fields coupled with gravity}},  \href{https://arxiv.org/abs/2007.02987}{{\tt
  2007.02987}}.

\bibitem{Bak:2018txn}
D.~Bak, C.~Kim and S.-H. Yi, \emph{{Bulk view of teleportation and traversable
  wormholes}}, \href{http://dx.doi.org/10.1007/JHEP08(2018)140}{\emph{JHEP}
  {\bf 08} (2018) 140}, [\href{https://arxiv.org/abs/1805.12349}{{\tt
  1805.12349}}].

\bibitem{Balasubramanian:2020hfs}
V.~Balasubramanian, A.~Kar, O.~Parrikar, G.~Sárosi and T.~Ugajin,
  \emph{{Geometric secret sharing in a model of Hawking radiation}},
  \href{https://arxiv.org/abs/2003.05448}{{\tt 2003.05448}}.

\bibitem{WI}
V.~Balasubramanian, A.~Kar and T.~Ugajin, \emph{{Islands in de Sitter space}},
  \href{https://arxiv.org/abs/2008.05275}{{\tt 2008.05275}}.

\bibitem{Chen:2020tes}
Y.~Chen, V.~Gorbenko and J.~Maldacena, \emph{{Bra-ket wormholes in
  gravitationally prepared states}},
  \href{https://arxiv.org/abs/2007.16091}{{\tt 2007.16091}}.

\bibitem{Hartman:2020khs}
T.~Hartman, Y.~Jiang and E.~Shaghoulian, \emph{{Islands in cosmology}},
  \href{https://arxiv.org/abs/2008.01022}{{\tt 2008.01022}}.

\bibitem{VanRaamsdonk:2020tlr}
M.~Van~Raamsdonk, \emph{{Comments on wormholes, ensembles, and cosmology}},
  \href{https://arxiv.org/abs/2008.02259}{{\tt 2008.02259}}.

\bibitem{Chen:2019uhq}
H.~Z. Chen, Z.~Fisher, J.~Hernandez, R.~C. Myers and S.-M. Ruan,
  \emph{{Information Flow in Black Hole Evaporation}},
  \href{https://arxiv.org/abs/1911.03402}{{\tt 1911.03402}}.

\bibitem{Chen:2020uac}
H.~Z. Chen, R.~C. Myers, D.~Neuenfeld, I.~A. Reyes and J.~Sandor,
  \emph{{Quantum Extremal Islands Made Easy, Part I: Entanglement on the
  Brane}},  \href{https://arxiv.org/abs/2006.04851}{{\tt 2006.04851}}.

\bibitem{Chen:2020jvn}
H.~Z. Chen, Z.~Fisher, J.~Hernandez, R.~C. Myers and S.-M. Ruan,
  \emph{{Evaporating Black Holes Coupled to a Thermal Bath}},
  \href{https://arxiv.org/abs/2007.11658}{{\tt 2007.11658}}.

\bibitem{Rozali:2019day}
M.~Rozali, J.~Sully, M.~Van~Raamsdonk, C.~Waddell and D.~Wakeham,
  \emph{{Information radiation in BCFT models of black holes}},
  \href{https://arxiv.org/abs/1910.12836}{{\tt 1910.12836}}.

\bibitem{Sully:2020pza}
J.~Sully, M.~Van~Raamsdonk and D.~Wakeham, \emph{{BCFT entanglement entropy at
  large central charge and the black hole interior}},
  \href{https://arxiv.org/abs/2004.13088}{{\tt 2004.13088}}.

\bibitem{Liu:2020gnp}
H.~Liu and S.~Vardhan, \emph{{A dynamical mechanism for the Page curve from
  quantum chaos}},  \href{https://arxiv.org/abs/2002.05734}{{\tt 2002.05734}}.

\bibitem{Liu:2020jsv}
H.~Liu and S.~Vardhan, \emph{{Entanglement entropies of equilibrated pure
  states in quantum many-body systems and gravity}},
  \href{https://arxiv.org/abs/2008.01089}{{\tt 2008.01089}}.

\bibitem{Hollowood:2020cou}
T.~J. Hollowood and S.~P. Kumar, \emph{{Islands and Page Curves for Evaporating
  Black Holes in JT Gravity}},  \href{https://arxiv.org/abs/2004.14944}{{\tt
  2004.14944}}.

\bibitem{Hollowood:2020kvk}
T.~J. Hollowood, S.~Prem~Kumar and A.~Legramandi, \emph{{Hawking Radiation
  Correlations of Evaporating Black Holes in JT Gravity}},
  \href{https://arxiv.org/abs/2007.04877}{{\tt 2007.04877}}.

\bibitem{Banks:2020zrt}
T.~Banks, \emph{{Microscopic Models of Linear Dilaton Gravity and Their
  Semi-classical Approximations}},
  \href{https://arxiv.org/abs/2005.09479}{{\tt 2005.09479}}.

\bibitem{Geng:2020qvw}
H.~Geng and A.~Karch, \emph{{Massive Islands}},
  \href{https://arxiv.org/abs/2006.02438}{{\tt 2006.02438}}.

\bibitem{Krishnan:2020fer}
C.~Krishnan, \emph{{Critical Islands}},
  \href{https://arxiv.org/abs/2007.06551}{{\tt 2007.06551}}.

\bibitem{Almheiri:2020cfm}
A.~Almheiri, T.~Hartman, J.~Maldacena, E.~Shaghoulian and A.~Tajdini,
  \emph{{The entropy of Hawking radiation}},
  \href{https://arxiv.org/abs/2006.06872}{{\tt 2006.06872}}.

\bibitem{Saad:2019lba}
P.~Saad, S.~H. Shenker and D.~Stanford, \emph{{JT gravity as a matrix
  integral}},  \href{https://arxiv.org/abs/1903.11115}{{\tt 1903.11115}}.

\bibitem{Marolf:2020xie}
D.~Marolf and H.~Maxfield, \emph{{Transcending the ensemble: baby universes,
  spacetime wormholes, and the order and disorder of black hole information}},
  \href{https://arxiv.org/abs/2002.08950}{{\tt 2002.08950}}.

\bibitem{Balasubramanian:2020jhl}
V.~Balasubramanian, A.~Kar, S.~F. Ross and T.~Ugajin, \emph{{Spin structures
  and baby universes}},  \href{https://arxiv.org/abs/2007.04333}{{\tt
  2007.04333}}.

\bibitem{Lewkowycz:2013nqa}
A.~Lewkowycz and J.~Maldacena, \emph{{Generalized gravitational entropy}},
  \href{http://dx.doi.org/10.1007/JHEP08(2013)090}{\emph{JHEP} {\bf 08} (2013)
  090}, [\href{https://arxiv.org/abs/1304.4926}{{\tt 1304.4926}}].

\bibitem{Asplund:2014coa}
C.~T. Asplund, A.~Bernamonti, F.~Galli and T.~Hartman, \emph{{Holographic
  Entanglement Entropy from 2d CFT: Heavy States and Local Quenches}},
  \href{http://dx.doi.org/10.1007/JHEP02(2015)171}{\emph{JHEP} {\bf 02} (2015)
  171}, [\href{https://arxiv.org/abs/1410.1392}{{\tt 1410.1392}}].

\bibitem{Sarosi:2016oks}
G.~Sárosi and T.~Ugajin, \emph{{Relative entropy of excited states in two
  dimensional conformal field theories}},
  \href{http://dx.doi.org/10.1007/JHEP07(2016)114}{\emph{JHEP} {\bf 07} (2016)
  114}, [\href{https://arxiv.org/abs/1603.03057}{{\tt 1603.03057}}].

\bibitem{Azeyanagi:2007qj}
T.~Azeyanagi, A.~Karch, T.~Takayanagi and E.~G. Thompson, \emph{{Holographic
  calculation of boundary entropy}},
  \href{http://dx.doi.org/10.1088/1126-6708/2008/03/054}{\emph{JHEP} {\bf 03}
  (2008) 054}, [\href{https://arxiv.org/abs/0712.1850}{{\tt 0712.1850}}].

\bibitem{Calabrese:2004eu}
P.~Calabrese and J.~L. Cardy, \emph{{Entanglement entropy and quantum field
  theory}}, \href{http://dx.doi.org/10.1088/1742-5468/2004/06/P06002}{\emph{J.
  Stat. Mech.} {\bf 0406} (2004) P06002},
  [\href{https://arxiv.org/abs/hep-th/0405152}{{\tt hep-th/0405152}}].

\bibitem{Maldacena:2016upp}
J.~Maldacena, D.~Stanford and Z.~Yang, \emph{{Conformal symmetry and its
  breaking in two dimensional Nearly Anti-de-Sitter space}},
  \href{http://dx.doi.org/10.1093/ptep/ptw124}{\emph{PTEP} {\bf 2016} (2016)
  12C104}, [\href{https://arxiv.org/abs/1606.01857}{{\tt 1606.01857}}].

\bibitem{Almheiri:2014cka}
A.~Almheiri and J.~Polchinski, \emph{{Models of AdS$_{2}$ backreaction and
  holography}}, \href{http://dx.doi.org/10.1007/JHEP11(2015)014}{\emph{JHEP}
  {\bf 11} (2015) 014}, [\href{https://arxiv.org/abs/1402.6334}{{\tt
  1402.6334}}].

\bibitem{Brown:2019rox}
A.~R. Brown, H.~Gharibyan, G.~Penington and L.~Susskind, \emph{{The Python's
  Lunch: geometric obstructions to decoding Hawking radiation}},
  \href{https://arxiv.org/abs/1912.00228}{{\tt 1912.00228}}.

\bibitem{Engelhardt:2020qpv}
N.~Engelhardt, S.~Fischetti and A.~Maloney, \emph{{Free Energy from Replica
  Wormholes}},  \href{https://arxiv.org/abs/2007.07444}{{\tt 2007.07444}}.

\bibitem{Verlinde:2020upt}
H.~Verlinde, \emph{{ER = EPR revisited: On the Entropy of an Einstein-Rosen
  Bridge}},  \href{https://arxiv.org/abs/2003.13117}{{\tt 2003.13117}}.

\bibitem{VanRaamsdonk:2010pw}
M.~Van~Raamsdonk, \emph{{Building up spacetime with quantum entanglement}},
  \href{http://dx.doi.org/10.1142/S0218271810018529}{\emph{Gen. Rel. Grav.}
  {\bf 42} (2010) 2323--2329}, [\href{https://arxiv.org/abs/1005.3035}{{\tt
  1005.3035}}].

\bibitem{Balasubramanian:2014gla}
V.~Balasubramanian, M.~Berkooz, S.~F. Ross and J.~Simon, \emph{{Black Holes,
  Entanglement and Random Matrices}},
  \href{http://dx.doi.org/10.1088/0264-9381/31/18/185009}{\emph{Class. Quant.
  Grav.} {\bf 31} (2014) 185009}, [\href{https://arxiv.org/abs/1404.6198}{{\tt
  1404.6198}}].

\bibitem{Takayanagi:2011zk}
T.~Takayanagi, \emph{{Holographic Dual of BCFT}},
  \href{http://dx.doi.org/10.1103/PhysRevLett.107.101602}{\emph{Phys. Rev.
  Lett.} {\bf 107} (2011) 101602}, [\href{https://arxiv.org/abs/1105.5165}{{\tt
  1105.5165}}].

\end{thebibliography}\endgroup

\end{document}